\documentclass[useAMS,usenatbib]{mn2e}

\newcommand{\kms}{ km s$^{-1}$\xspace}
\newcommand{\ang}{\mbox{\AA}\xspace}
\newcommand{\nhI}{N$_{\rm H \ I}$\xspace}

\newcommand{\za}{z$_{abs}$}
\newcommand{\lala}{$\lambda\lambda$\xspace}
\newcommand\Lya{Lyman-$\alpha$\xspace}

\usepackage{graphics}
\usepackage{graphicx}
\usepackage{colordvi}
\usepackage{epsfig}
\usepackage{lscape}
\usepackage{rotating}
\usepackage{xspace}
\usepackage{float}
\begin{document}

\title[Abundances of Sub-DLAs at z $\la$ 1.5.]{New MIKE Observations of $z < 1.5$ Sub-Damped Lyman-$\alpha$ Systems}
\author[J. D. Meiring, V.P. Kulkarni et al.]{Joseph D. Meiring$^{1}$\thanks{Current Address: Department of Physics and Astronomy, University of Louisville, Louisville Ky 40292}, 
Varsha P. Kulkarni$^{1}$, 
James T. Lauroesch$^{2}$,
Celine P\'eroux$^{3}$,
\newauthor  Pushpa Khare$^{4}$, $\&$ Donald G. York$^{5,6}$ \\
$^{1}$Department of Physics and Astronomy, University of South Carolina, Columbia, SC 29208, USA \\
$^{2}$Department of Physics and Astronomy, University of Louisville, Louisville, Ky 40292 USA\\
$^{3}$Laboratoire d'Astrophysique de Marseille, OAMP, Universite Aix-Marseill\`e $\&$ CNRS, 13388 Marseille cedex 13, France \\
$^{4}$Department of Physics, Utkal University, Bhubaneswar, 751004, India \\
$^{5}$Department of Astronomy and Astrophysics, University of Chicago, Chicago, IL 60637, USA \\ 
$^{6}$Enrico Fermi Institute, University of Chicago, Chicago, IL 60637, USA \\ }
%$^{7}$Department of Astronomy, Columbia University, New York, NY 10027, USA\\}

\date{Accepted ... Received ...; in original form ...}

\pagerange{\pageref{firstpage}--\pageref{lastpage}} \pubyear{}

\maketitle

\label{firstpage}

\begin{abstract}

The Damped and sub-Damped \Lya (DLA and sub-DLA) systems seen in the spectra of QSOs offer a unique way to study the interstellar medium of high
redshift galaxies. In this paper we report on new abundance determinations in a sample of 10 new systems, nine of the lesser studied 
sub-DLAs and one DLA, along the line of sight to seven QSOs from spectra taken with the MIKE spectrograph. 
Lines of Mg I, Mg II, Al II, Al III, Ca II, Mn II, Fe II, and Zn II were detected. 
Here, we give the column densities and equivalent widths of the observed absorption lines, as well as the abundances determined for these systems.
Zn, a relatively undepleted element in the local interstellar medium (ISM) is detected in one system with a high metallicity of [Zn/H]=+0.27$\pm$0.18. 
In one other system, a high abundance based on the more depleted element Fe is seen with [Fe/H]=$-0.37\pm0.13$, although Zn is not detected. 
The \nhI-weighted mean metallicity of these sub-DLA systems based on Fe is $\langle$[Fe/H]$\rangle$=$-0.76\pm0.11$, 
nearly $\sim$0.7 dex higher (a factor of 5) than what is seen in DLAs
in this redshift range. The relative abundance of [Mn/Fe] is also investigated. A clear trend is visible
for these systems as well as systems from the literature, 
with [Mn/Fe] increasing with increasing metallicity in good agreement with with Milky Way stellar abundances.

\end{abstract}

\begin{keywords}
{Quasars:} absorption lines-{ISM:} abundances
\end{keywords}

\section{Introduction}

Quasar absorption line systems with strong Lyman-$\alpha$ lines are often divided into two classes: Damped Lyman-$\alpha$ (DLAs, log \nhI $\ge$ 20.3) and 
sub-Damped Lyman-$\alpha$ (sub-DLA 19 $\la$ log \nhI $<$ 20.3, \citealt{Per01}) which contain a major fraction of the neutral gas in the Universe, 
while the majority of the baryons are thought to lie in the highly ionized and diffuse Lyman-$\alpha$ forest clouds with log \nhI $\la$ 15 in intergalactic space
\citep{Petit93, Dan08}. The lower threshold of log \nhI=20.3 for classification of DLAs stems from previous 21 cm emission studies of nearby spirals, where 
the sensitivity limited column density of log \nhI$\sim$20.3 was seen to lie near the Holmberg radius (R$_{26.5}$) of the galaxy \citet{Bos81}. Nonetheless, the damping wings 
which can be used to accurately measure \nhI in these systems do begin the sub-DLA regime of log \nhI$\ga$19.0.
With their high gas content, the DLA and sub-DLA systems are believed to be associated directly with galaxies at all redshifts in which they are seen. 

Among the many elements often detected in QSO absorber systems including C, N, O, Mg, Si, S, Ca, Ti, Cr, Mn, Fe, Ni, and Zn,  Zn is 
the preferred tracer of the gas-phase metallicity as it is relatively undepleted in the Galactic ISM, especially 
when the fraction of H in molecular form is low, as is the case in most DLAs. Zn also tracks the Fe abundance in Galactic stars (e.g., \citet{Niss04}), 
and the lines of Zn II \lala 2026,2062 are relatively weak and typically unsaturated.
These lines can be covered with ground based spectroscopy over a wide range of redshifts, 0.65 $\la$ z $\la$ 3.5, which covers a large portion ($\sim 45\%$) of the 
history of the Universe. Abundances of refractory elements such as Cr and Fe relative to Zn also give a measure of the amount of dust depletion \citep{York06}.
Abundance ratios such as [Si/Fe], [O/Fe] and [Mn/Fe] shed light on the enrichment from the different types of supernovae, as the $\alpha$-capture elements 
Si and O for example are produces mainly in Type II explosions while the iron peak elements are produced mainly by Type Ia supernovae. 

In previous studies, DLA systems have been the preferred systems for chemical abundance investigation owing to their high gas content \citep{PW02, Kul05, Mei06}. 
Most DLAs however have been found
to be metal poor, typically far below the solar level and below where models predict the mean metallicity should be at the corresponding redshifts at which they are seen
(e.g., \citet{Kul05} and references therein). The sub-DLA systems have until recently been largely ignored, with their contribution to the overall metal budget 
unknown. Evidence for the possibility of a non-negligible contribution from sub-DLAs to the metal budget 
came from \citet{Per03a}, who noted that based on Fe II lines the sub-DLA systems have faster evolution of the 
Fe abundance and higher abundances on average than DLA systems. This has also been validated by \citet{Kul07}.

Galactic chemical evolution is a slow process, and long timescales must be examined to search for the signs of the gradual chemical
enrichment that models predict \citep{Cen03, PFH99}.  
Although redshifts $z < 1.5$ span ~70$\%$ of the age of the Universe (using a concordance cosmology of $\Omega_{m}=0.3, \Omega_{\Lambda}=0.7$, H$_{0}$=70 \kms Mpc$^{-1}$), 
few observations have been made of $z < 1.5$ sub-DLAs due to the lack of spectrographs with enough sensitivity in short wavelengths and  
the paucity of known sub-DLAs in this redshift range. 
With such a large fraction of the age of the Universe covered in the redshift regime,
it is clearly important for understanding the nature of sub-DLA systems and galactic chemical evolution as well.
 
We have greatly increased the sample of sub-DLAs in past several years with our VLT UVES and Magellan-II MIKE spectra.
In this paper,  we report on ten new systems observed with the 
MIKE spectrograph on the Magellan-II Clay telescope. The structure of this paper is as follows: In $\S$ \ref{Sec:Obs}, 
we discuss details of our observations and data reduction techniques. 
$\S$ \ref{Sec:Objects} gives details  
on the individual objects in these new observations. $\S$ \ref{Sec:MnFe} investigates the ratio of [Mn/Fe] in QSO absorbers. 
In $\S$ \ref{Sec:Abund} we discuss the abundances of these absorbers
and we give a brief discussion.  We also provide an appendix at the end of this paper, showing plots of the 
UV spectra with the fits to the Lyman-$\alpha$ lines, and tables containing the fit parameters for the individual systems.

\section{Observations and Data Reduction } \label{Sec:Obs}
The observations presented here were made with the 6.5m Magellan-II Clay telescope and the Magellan Inamori Kyocera Echelle (MIKE) spectrograph \citep{Bern03} in 2007 Sep. 
This is a double sided spectrograph with both a blue and a red camera, providing for 
simultaneous wavelength coverage from $\sim$3340 \ang to $\sim$9400 \ang. Targets were observed in 
 multiple exposures of 1800 to 2700 sec each to minimize cosmic ray defects. The seeing was typically 
 $<$ 1$\arcsec$, averaging $\sim$ 0.7$\arcsec$. All of the target QSOs were observed 
 with the 1$\arcsec$x5$\arcsec$ slit and the spectra were binned 2x3 (spatial by spectral) during readout. 
The resolving power of the MIKE spectrograph is $\sim$19,000 and $\sim$25,000 on the red and blue sides respectively with a 1$\arcsec$ slit.
Table 1 gives a summary of the observations. 

These spectra were reduced using the MIKE pipeline reduction code in IDL developed by S. Burles, J. X. Prochaska, 
and R. Bernstein. Wavelengths were calibrated using a Th-Ar comparison lamp taken after each exposure. 
The data were first bias subtracted from the overscan region and flat-fielded. The data  were then sky-subtracted and the spectral orders were extracted
using the traces from flat field images. These extracted spectra were then corrected for heliocentric velocities and converted
to vacuum wavelengths. Each individual order was then combined in IRAF using rejection parameters to reduce the effects of cosmic rays.
These combined spectra were then normalized using a
polynomial, typically of order five or less, or spline function to fit the continuum.

Our new observations consists of 10 absorbers, nine sub-DLAs and one DLA \za $\la$ 1.5. 
See \citet{Mei08} for a discussion of our selection criteria. Throughout this paper the QSO names are given in J2000 coordinates, 
except in Table 1 where the original name, based on J1950 coordinates is also given if applicable \citep{HB87}.

\setlength{\tabcolsep}{4pt}
\setcounter{table}{1}
\begin{table*}
\begin{tabular}{llccccccccccc}
\hline
\hline
 QSO		 	&	 Original or SDSS ID	 	&	 	RA		 	&	 	Dec		 	&	m$_{V}$ or m$_{g}$ 	&	 $z_{em}$	 	&	 $z_{abs}$	 	&	\nhI				&	 	Exposure Time	 \\
 J2000		 	&	 			 	&	 			 	&	 			 	&	 		 	&	 		 	&	 		 	&	 	cm$^{-2}$ 	 	&	 	Sec		  \\
\hline																					 
Q0005+0524	 	&	 Q0002+051		 	&	       00:05:20.21	 	&	        +05:24:10.8	 	&	  	16.9$^{a}$ 	&	  	1.899	 	&	  	0.8514	 	&	  	19.08$\pm$0.04	 	&	  	3600		 \\
Q0012$-$0122		&	Q0009-016			&	00:12:10.89			&	$-$01:22:07.5			&	18.1$^{a}$		&	1.998			&		1.3862		&	20.26$\pm$0.03			&		5400		 \\
Q0021+0104		&	SDSS J002127.88+010420.2	&	00:21:27.88			&	+01:04:20.1			&	18.6$^{b}$		&	1.829			&	1.3259			&	20.04$\pm$0.11			&		8100		\\
$\cdots$		&	 $\cdots$		 	&	 $\cdots$			&	 $\cdots$		 	&	 $\cdots$		&	 $\cdots$		&	1.5756			&	20.48$\pm$0.15			&		$\cdots$	 \\
Q0427$-$1302		&	Q0424-131			&	04:27:07.32			&	$-$13:02:53.6			&	17.5$^{a}$		&	2.166			&	1.4080			&	19.04$\pm$0.04			&		12300		 \\
Q1631+1156		&	Q1629+120			&	16:31:45.24			&	+11:56:02.9			&	18.6$^{b}$		&	1.792			&	0.9004			&	19.70$\pm$0.04			&		13500		 \\
Q2051+1950		&	Q2048+196			&	20:51:45.87			&	+19:50:06.3			&	18.5$^{a}$		&	2.367			&	1.1157			&	20.00$\pm$0.15			&		13500		 \\
Q2352$-$0028		&	SDSS J235253.51-002850.4	&	23:52:53.51			&	$-$00:28:51.3			&	18.5$^{b}$		&	1.628			&	0.8730			&	19.18$\pm$0.09			&		13500		 \\
$\cdots$	 	&	 $\cdots$	 	 	&	 $\cdots$	 	 	&	 $\cdots$	 	 	&	 $\cdots$	  	&	 $\cdots$	 	&	1.0318			&	19.81$\pm$0.13			&		$\cdots$	 \\
$\cdots$	 	&	 $\cdots$	 	 	&	 $\cdots$	 	 	&	 $\cdots$	 	 	&	 $\cdots$	  	&	 $\cdots$	 	&	1.2467			&	19.60$\pm$0.24			&		$\cdots$	 \\
\hline
\end{tabular}
\textbf{Table 1:} Summary of Observations. $^{a}$m$_{V}$, $^{b}$m$_{g}$   \label{Tab1} 
\end{table*}

%%%%%%%%%%%%%%%%%%%%%%%%%%%%%%%%%%%%%%%%%%%%%%%%%%%%%%%%%%%%%%%%%%%%%%%%%%%%%%%%%%%%%%%%%%%%%%%%%%%%%%%%%%%%%%%%%%5
%%%%%%%%%%%%%%%%%%%%%%%%%%%%%%%%%%%%%%%%%%%%%%%%%%%%%%%%%%%%%%%%%%%%%%%%%%%%%%%%%%%%%%%%%%%%%%%%%%%%%%%%%%%%%%%%%%%%55

\section{Determination of Column Densities } \label{Sec:CD}
    Column densities were determined from profile fitting with the package FITS6P \citep{Wel91}, which has
evolved from the code by \citet{VM77}. FITS6P iteratively minimizes the $\chi^{2}$ value between the data and
a theoretical Voigt profile that is convolved with the instrumental profile. 
The profile fit used multiple components, tailored to the individual system. For the central, core components, the effective Doppler parameters
($b_{eff}$) and radial velocities were determined from the weak and unsaturated lines, typically the Mg I $\lambda$ 2852 line. 
For the weaker components at higher radial velocities the $b_{eff}$ and component velocity values were determined from stronger transitions such as the 
Fe II \lala 2344, 2382 lines and the Mg II \lala 2796, 2803 lines. A set of $b_{eff}$ and $v$ values were thus
determined that reasonably fit all of the lines observed in the system. 
The same $b_{eff}$ values were used for all the species. The atomic data used in line identification and profile fitting are from \citet{Morton03}.

	 In general, if a multiplet was observed, the lines were fit simultaneously until convergence. 
For all of the systems, the Fe II \lala 2344, 2374, 2382 lines were fit simultaneously to determine a set of column
densities that fit the spectra reasonably well. Similarly, the Mg II \lala 2796, 2803 lines were also fit together. 
Significant saturation of the Mg II \lala 2796, 2803 and Al II $\lambda$ 1670 lines allowed for only lower limits to be placed on the column densities 
for these species. The Zn II $\lambda$ 2026.137 line is blended with a line of Mg I $\lambda$ 2026.477. The Mg I contribution to the line was estimated using the
Mg I $\lambda$ 2852 line, for which f$\lambda$ is $\sim$32 times that of the  Mg I $\lambda$ 2026 line. 
The Zn II components were then allowed to vary while the Mg I components were held fixed. 
N$_{\rm Cr \ II}$ was determined by simultaneously fitting the Cr II $\lambda$ 2056 line and the blended Cr II + Zn II $\lambda$ 2062 line, where the 
contribution from Zn II was estimated from the Zn II + Mg I $\lambda$ 2026 line. See also \citet{Kh04} and \citet{Mei07, Mei08} for a discussion of the profile fitting scheme.
We adopt the standard notation:

\begin{equation}
\rm{[X/H] = log(N_{\rm X}/N_{\rm H \ I}) - log (X/H)_{\sun}}
\end{equation}

	Solar system abundances have been adopted from \citet{Lodd03}. As the \Lya lines from which we can determine \nhI all lie in the UV, even when redshifted, 
space based UV spectra are necessary. Neutral Hydrogen column densities were determined from archived HST STIS and FOS spectra available from the HST archives. These systems 
have previously had \nhI determined in \citet{Rao06}, although the plots for the sub-DLAs are not published. We give in the Appendix the plots of the \Lya lines for these
systems with the fits overlaid. In general, the fits from \citet{Rao06} and our determinations agree quite well. However, for Q2051+1960 we have determined a different value of
\nhI=20.00$\pm$0.15. In all other cases, we have used the values from \citet{Rao06}. Rest frame equivalent widths for the lines are given in Table 2,
with 3$\sigma$ upper limits based on the photon noise and continuum level given in cases without a detection.

\setlength{\tabcolsep}{4pt}
\setcounter{table}{2}
\begin{sidewaystable*}
\begin{minipage}{200mm}
\textbf{Table 2:} Rest frame equivalent widths and 1$\sigma$ errors of the observed absorption lines in m\ang. 
\begin{center}
\begin{tabular}{llrrrrrrrrrrrrrrrrrrr}
\hline
\hline
QSO			&		z$_{abs}$	&		Mg I		&		Mg II	 		&		Mg II 			&		Al II 	 		&		Al III 	 		&		Al III 		 	&		Si II 		 	&		Si II 		 	&		Ca II 		 	&		Ca II			&		Cr II 			\\
			&				&		2852		&		2796			&		2803			&		1670			&		1854			&		1862			&		1526			&		1808			&		3933			&		3969			&		2056		 	\\
\hline
Q0005+0524		&		0.8514		&		201$\pm$19	&		1104$\pm$32		&		976$\pm$29		&		-			&		171$\pm$20		&		104$\pm$17		&			-		&		-			&		-			&		-			&		$<$3	\\
Q0012$-$0122		&		1.3862		&		69$\pm$9	&		1115$\pm$14		&		891$\pm$15		&		290$\pm$12		&		94$\pm$9		&		44$\pm$9		&		304$\pm$4		&		-			&		-			&		-			&		$<$3	\\
Q0021+0104A		&		1.3259		&		206$\pm$31	&		2727$\pm$28		&		2370$\pm$37		&		1038$\pm$40		&		-			&		-			&		880$\pm$47		&		$<$12			&		-			&		-			&		$<$6	\\
Q0021+0104B		&		1.5756		&		459$\pm$61	&		3010$\pm$60		&		2468$\pm$71		&		-			&		$<$9			&		$<$6			&		1106$\pm$54		&		$<$5			&		-			&		-			&		$<$15	\\
Q0427$-$1302		&		1.4080		&		-		&		316$\pm$8		&		225$\pm$7		&		61$\pm$5		&		$<$3			&		-			&		-			&		-			&		-			&		-			&		$<$3	\\
Q1631+1156		&		0.9004		&		257$\pm$41	&		1081$\pm$29		&		908$\pm$38		&		-			&		-			&		-			&		-			&		-			&		133$\pm$34		&		-			&		$<$17	\\
Q2051+1950		&		1.1157		&		454$\pm$23	&		1528$\pm$13		&		1448$\pm$11		&		747$\pm$37		&		392$\pm$18		&		-			&		-			&		105$\pm$33		&		300$\pm$27		&		-			&		44$\pm$11	\\
Q2352$-$0028A		&		0.8730		&		85$\pm$28	&		1403$\pm$26		&		1075$\pm$36		&		-			&		-			&		-			&		-			&		-			&		$<$9			&		-			&		$<$9	\\
Q2352$-$0028B		&		1.0318		&		445$\pm$29	&		2069$\pm$18		&		1984$\pm$16		&		-			&		809$\pm$40$^{c}$	&		189$\pm$22		&		-			&		170$\pm$24		&		-			&		-			&		$<$6	\\
Q2352$-$0028C		&		1.2467		&		247$\pm$46	&		2975$\pm$28		&		2342$\pm$55		&		711$\pm$28		&		374$\pm$24		&		229$\pm$22		&		-			&		$<$5			&		-			&		-			&		$<$6	\\
\hline
QSO			&		z$_{abs}$	&		Mn II 		&		Mn II 			&		Mn II 	 		&		Fe II 	 		&		Fe II 	 		&		Fe II		 	&		Fe II 		 	&		Fe II 		 	&		Fe II 		 	&		Zn II$^{a}$ 		&		Zn II$^{b}$		\\
			&				&		2576		&		2594			&		2606			&		2260			&		2344			&		2374			&		2382			&		2586			&		2600			&		2026			&		2062			\\
\hline
Q0005+0524		&		0.8514		&		$<$3		&		$<$2			&		$<$3			&		$<$3			&		233$\pm$10		&		98$\pm$12		&		382$\pm$8		&		179$\pm$10		&		417$\pm$12		&		$<$3			&		$<$4	\\
Q0012$-$0122		&		1.3862		&		$<$6		&		$<$7			&		$<$7			&		$<$10			&		330$\pm$8		&		187$\pm$6		&		460$\pm$18		&		-			&		460$\pm$15		&		$<$3			&		$<$4	\\
Q0021+0104A		&		1.3259		&		$<$18		&		-			&		-			&		$<$16			&		1261$\pm$47		&		571$\pm$57		&		1829$\pm$42		&		1106$\pm$54		&		1836$\pm$59		&		$<$6			&		$<$7	\\
Q0021+0104B		&		1.5756		&	$<$10			&	$<$13				&	$<$14				&		$<$22			&		1181$\pm$55		&		518$\pm$49		&		1873$\pm$69		&		982$\pm$41		&		1766$\pm$31		&		$<$16			&		$<$16	\\
Q0427$-$1302		&		1.4080		&		$<$10		&	$<$10				&		-			&		$<$6			&		90$\pm$7		&		38$\pm$6		&		168$\pm$6		&		85$\pm$5		&		173$\pm$5		&		$<$3			&		$<$3	\\
Q1631+1156		&		0.9004		&		$<$74		&		-			&		-			&		$<$7			&		-			&		-			&		617$\pm$21		&		400$\pm$28		&		624$\pm$78		&		$<$28			&		$<$25	\\
Q2051+1950		&		1.1157		&		312$\pm$29	&		269$\pm$39		&		152$\pm$21		&		98$\pm$13		&		1118$\pm$15		&		831$\pm$32		&		1211$\pm$42		&		1154$\pm$16		&		1316$\pm$26		&		152$\pm$27		&		108$\pm$25	\\
Q2352$-$0028A		&		0.8730		&		-		&		$<$5			&		-			&		$<$5			&		181$\pm$32		&		$<$5			&		362$\pm$15		&		78$\pm$12		&		266$\pm$12		&		$<$8			&		$<$15	\\
Q2352$-$0028B		&		1.0318		&		-		&		$<$12			&		-			&		69$\pm$19		&		1362$\pm$9		&		761$\pm$20		&		1537$\pm$12		&		1304$\pm$16		&		1553$\pm$15		&		$<$18			&		$<$15 	\\
Q2352$-$0028C		&		1.2467		&		$<$11		&		-			&		-			&		$<$13			&		637$\pm$40		&		216$\pm$51		&		1151$\pm$50		&		508$\pm$56		&		936$\pm$37		&		$<$6			&		$<$5 	\\
\hline
\end{tabular}	
\end{center}																					
$^{a}$This line is a blend with Mg I $\lambda$ 2026, although the Mg I contribution is judged to be insignificant in all cases.\\																							
$^{b}$As this line is blended with the Cr II $\lambda$ 2062 line, this value represents the total EW of the line.\\	
$^{c}$Blended with another feature.\\																						
\end{minipage}
\end{sidewaystable*}

%%%%%%%%%%%%%%%%%%%%%%%%%%%%%%%%%%%%%%%%%%%%%%%%%%%%%%%%%%%%%%%%%%%%%%%%%%%%%%%%%%%%%%%%%%%%%%%%
%%%%%%%%%%%%%%%%%%%%%%%%%%%%%%%%%%%%%%%%%%%%%%%%%%%%%%%%%%%%%%%%%%%%%%%%%%%%%%%%%%%%%%%%%%%%%%%%

\section{Notes on Individual Objects } \label{Sec:Objects}

\subsubsection{Q0005+0524 ($z_{em}=1.899$)}
\textbf{(System A:\za=0.8514):} This fairly bright QSO has a weak sub-DLA system with log \nhI=19.08 in the spectrum \citep{Rao06}. Eight components were needed to fit the 
observed profiles. Lines of Mg I $\lambda$ 2852, Mg II \lala 2796, 2803, Al III \lala 1854, 1862, and Fe II \lala 2344, 2374, 2382, 2586, 2600 were detected. No Zn II \lala 2026, 2052 lines were
detected with S/N$\sim$50 in the region. An upper limit of [Zn/H]$<-$0.47 was placed based on the noise in the region. Based on Fe, the metallicity for this system is
[Fe/H]=$-$0.76$\pm$0.04. With the relatively low value of Fe II/ Al III = +0.68$\pm$0.04, there could be significant ionisation in the system. 
Velocity plots of several lines are shown in Figure \ref{Fig:Q0005}.

\subsubsection{Q0012$-$0122 ($z_{em}=1.998$)}
\begin{figure*} 
\begin{minipage}{7in}
\begin{center}
\includegraphics[angle=90, width=5.5in]{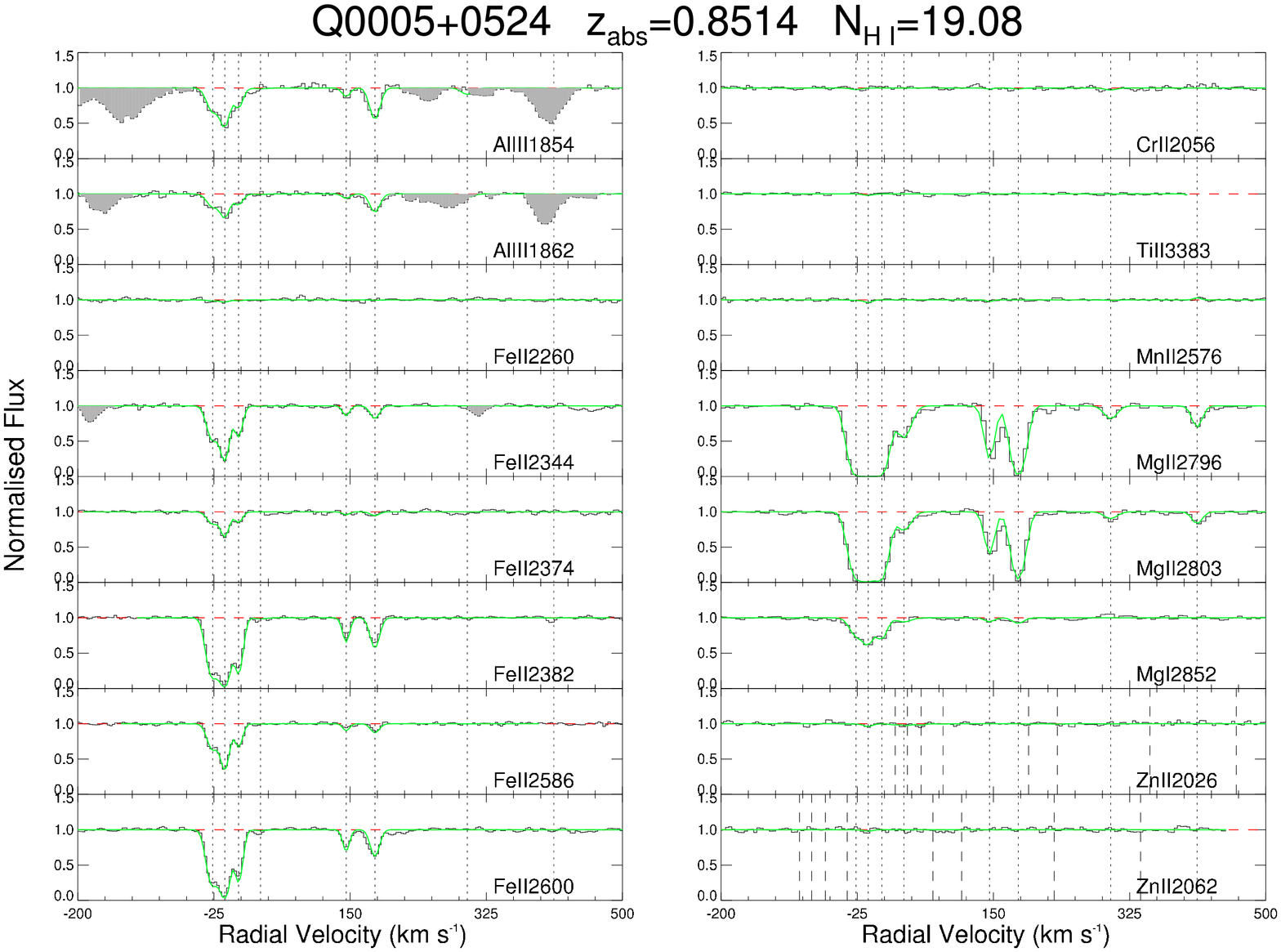}
\caption{Velocity plots for Q0005+0524. The solid green line indicates the theoretical profile fit to the spectrum, and the dashed red line is the continuum level.
The vertical dotted lines indicate the positions of the components that were used in the fit. In the cases of the Zn II $\lambda\lambda$ 2026,2062
lines, the long dashed vertical lines indicate the positions of the components for Mg I (former case), and Cr II (latter case). Areas shaded in gray are due to 
interloping absorption features or cosmetic defects. \label{Fig:Q0005}}
\includegraphics[angle=90, width=5.5in]{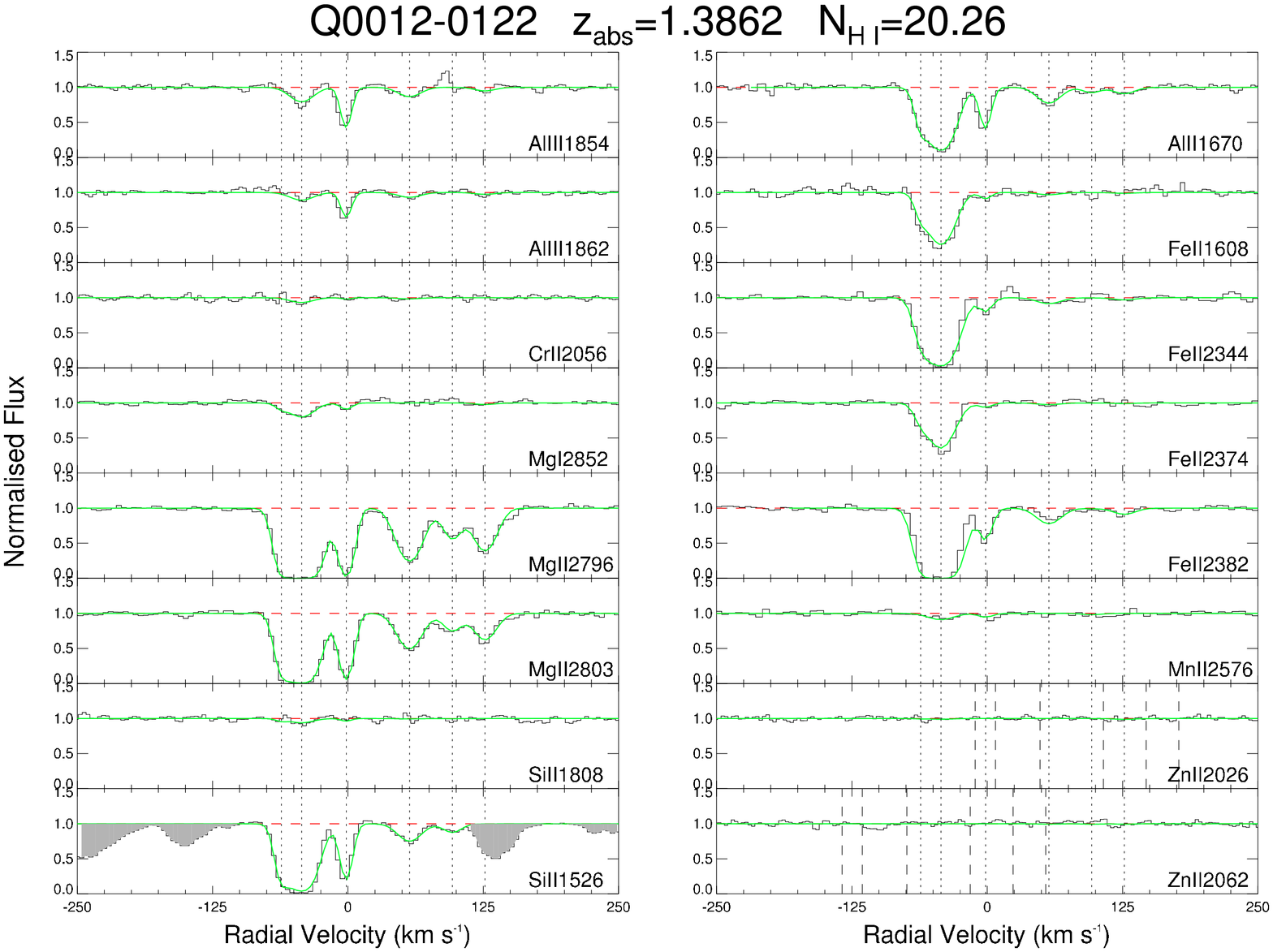}
\end{center}
\caption{ Same as Figure  \ref{Fig:Q0005} but for Q0012-0122 \label{Fig:Q0012} }
\end{minipage}
\end{figure*}

\textbf{(System A:\za=1.3862):} This system is a sub-DLA with log \nhI=20.26 \citep{Rao06}.  Lines of Mg I $\lambda$ 2852, Mg II \lala 2796, 2803, Al II $\lambda$ 1670, Al III \lala 1854, 1862, Si II $\lambda$ 1526 and 
Fe II \lala 2344, 2374, 2382, 2586, 2600 were all detected. This system shows minimal  $\alpha$ enhancement with [Si/Fe]=+0.12$\pm$0.08. 
Detections of both the Al III and Al II lines in this system show that Al III/Al II$<-0.19$. 
This system appears to be fairly metal poor. No Zn II \lala 2026, 2062 lines were detected and an 
upper limit of [Zn/H]$<-1.34$ was determined for this system. Based on Fe, the metallicity is [Fe/H]=$-1.49\pm0.02$. 
Velocity plots of several lines are shown in Figure \ref{Fig:Q0012}.

\begin{figure*} 
\begin{minipage}{7in}
\begin{center}
\includegraphics[angle=90, width=5.5in]{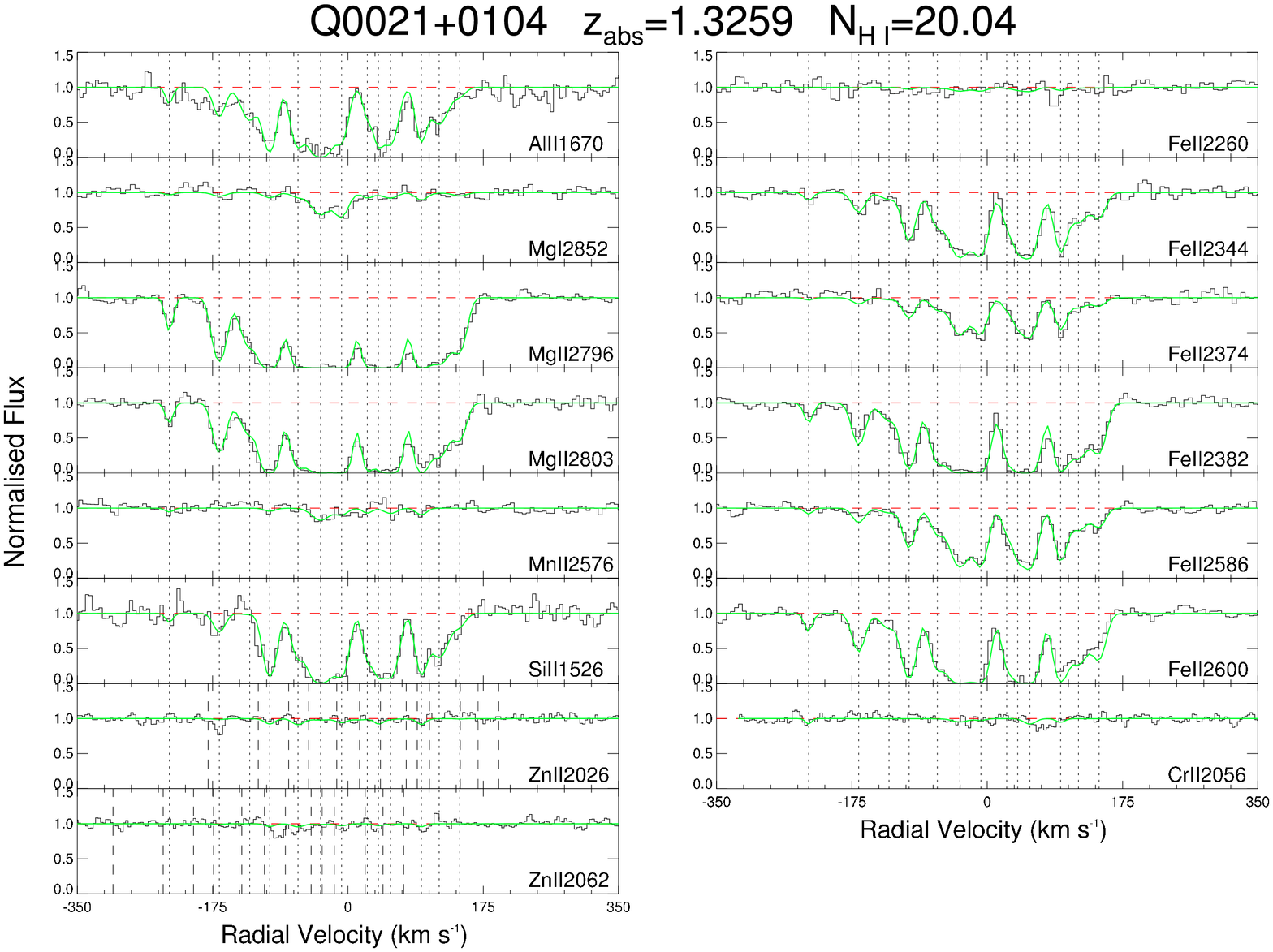}
\caption{Same as Figure  \ref{Fig:Q0005} but for Q0021+0104A \label{Fig:Q0021A} }
\includegraphics[angle=90, width=5.5in]{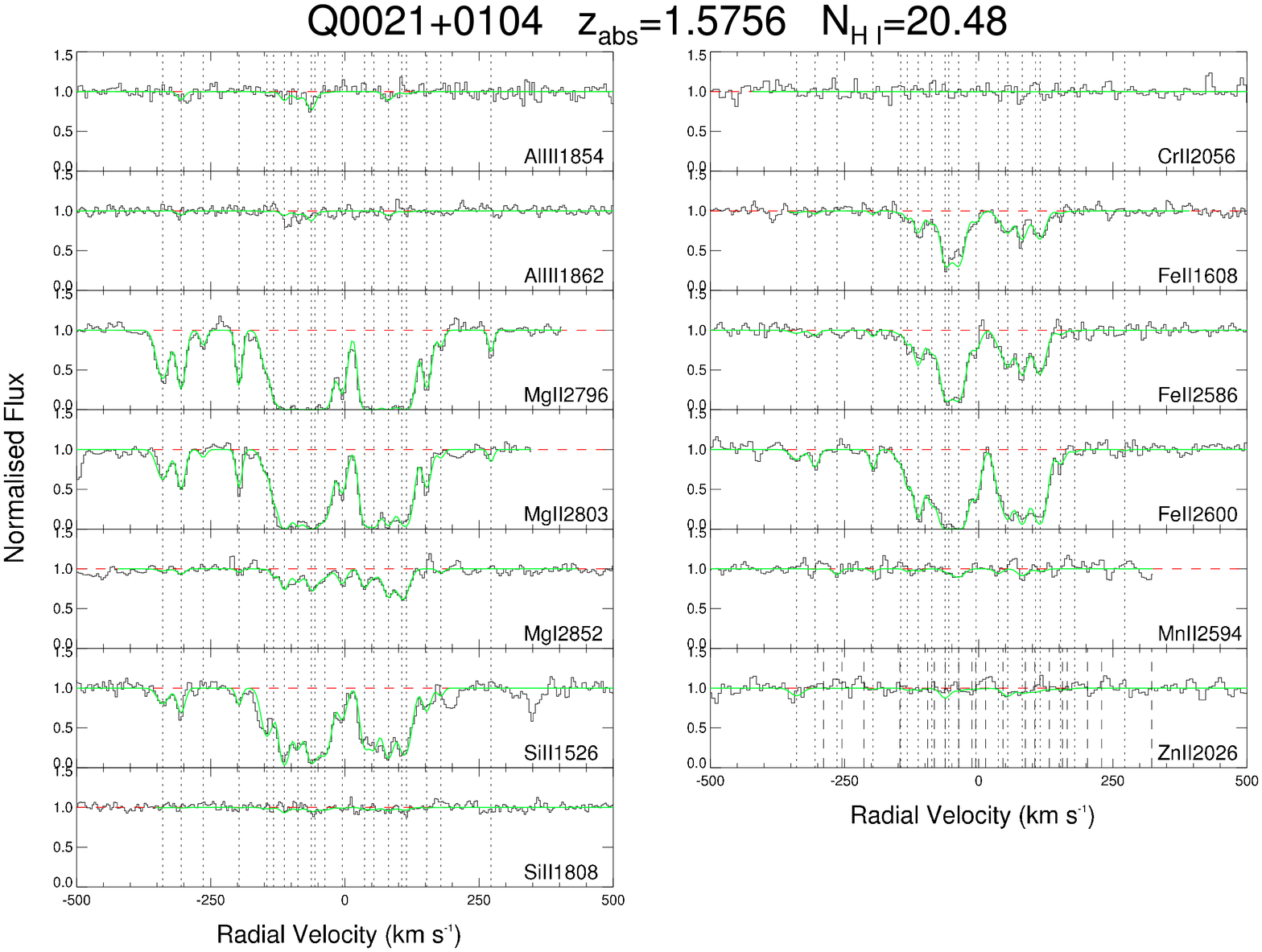}
\end{center}
\caption{ Same as Figure  \ref{Fig:Q0005} but for Q0021+0104B \label{Fig:Q0021B} }
\end{minipage}
\end{figure*}

\begin{figure*} 
\begin{minipage}{7in}
\begin{center}
\includegraphics[angle=90, width=5.5in]{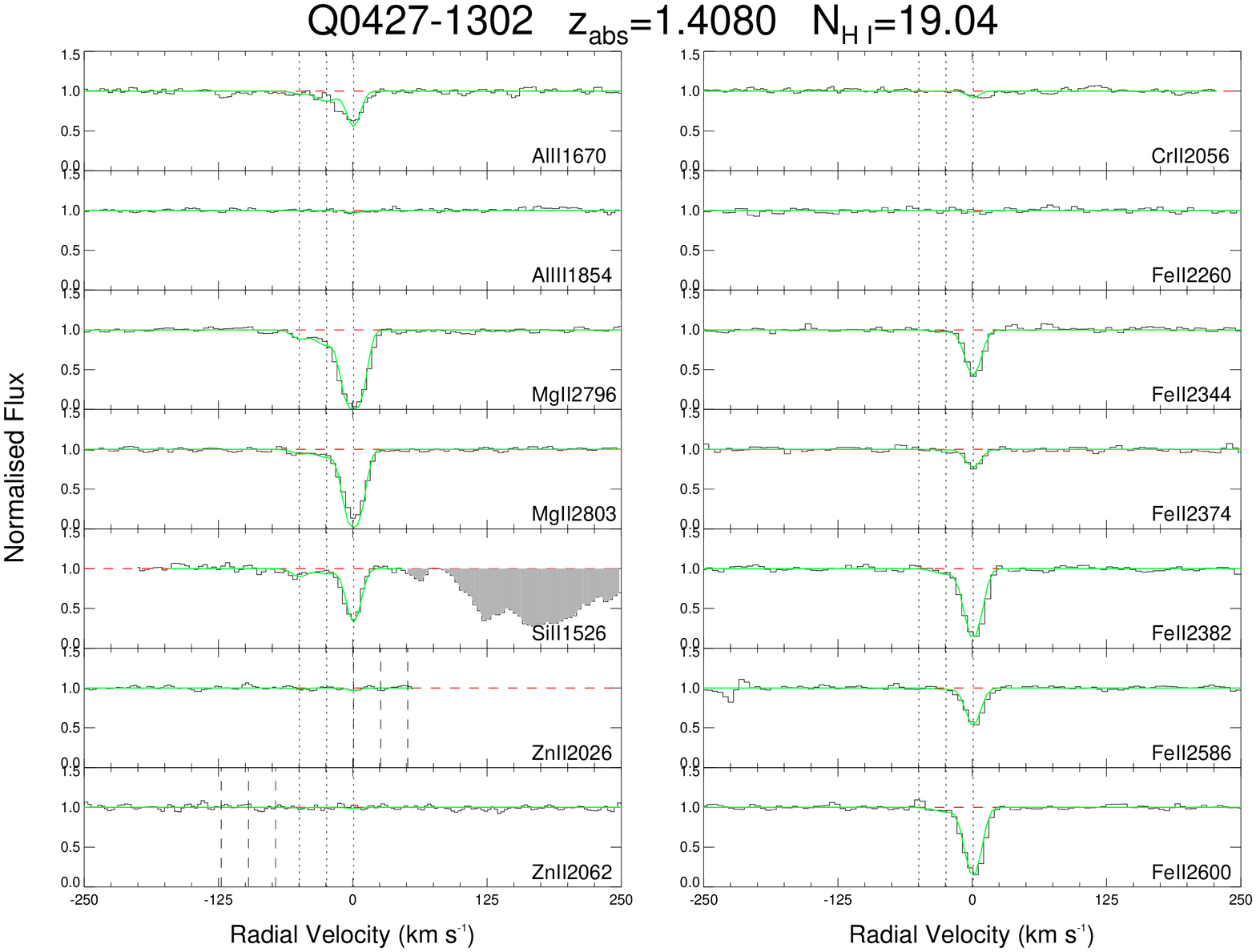}
\caption{Same as Figure  \ref{Fig:Q0005} but for Q0427-1302 \label{Fig:Q0427}}
\includegraphics[angle=90, width=5.5in]{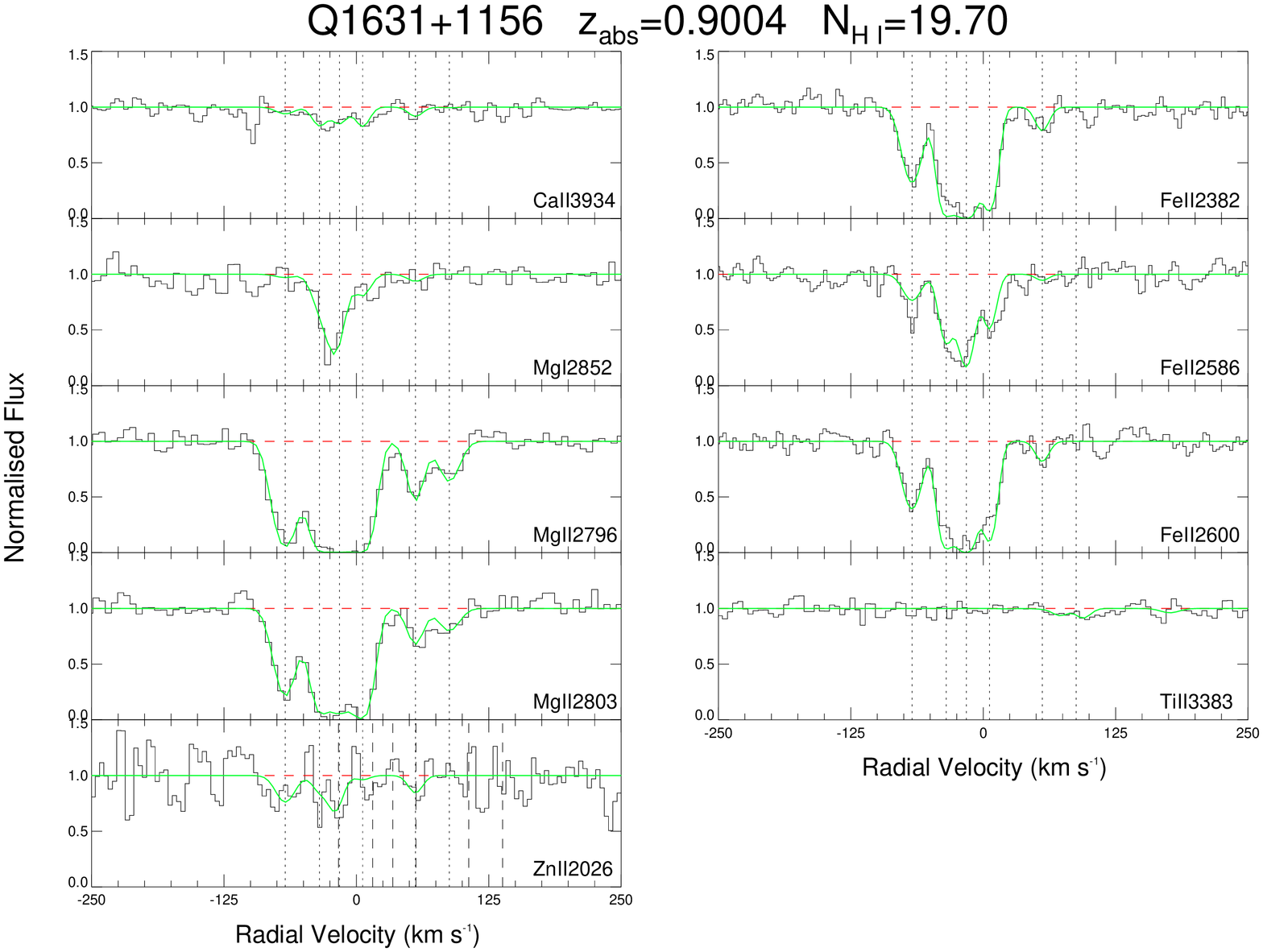}
\end{center}
\caption{ Same as Figure  \ref{Fig:Q0005} but for Q1631+1156 \label{Fig:Q1631} }
\end{minipage}
\end{figure*}

\begin{figure*} 
\begin{minipage}{7in}
\begin{center}
\includegraphics[angle=90, width=5.5in]{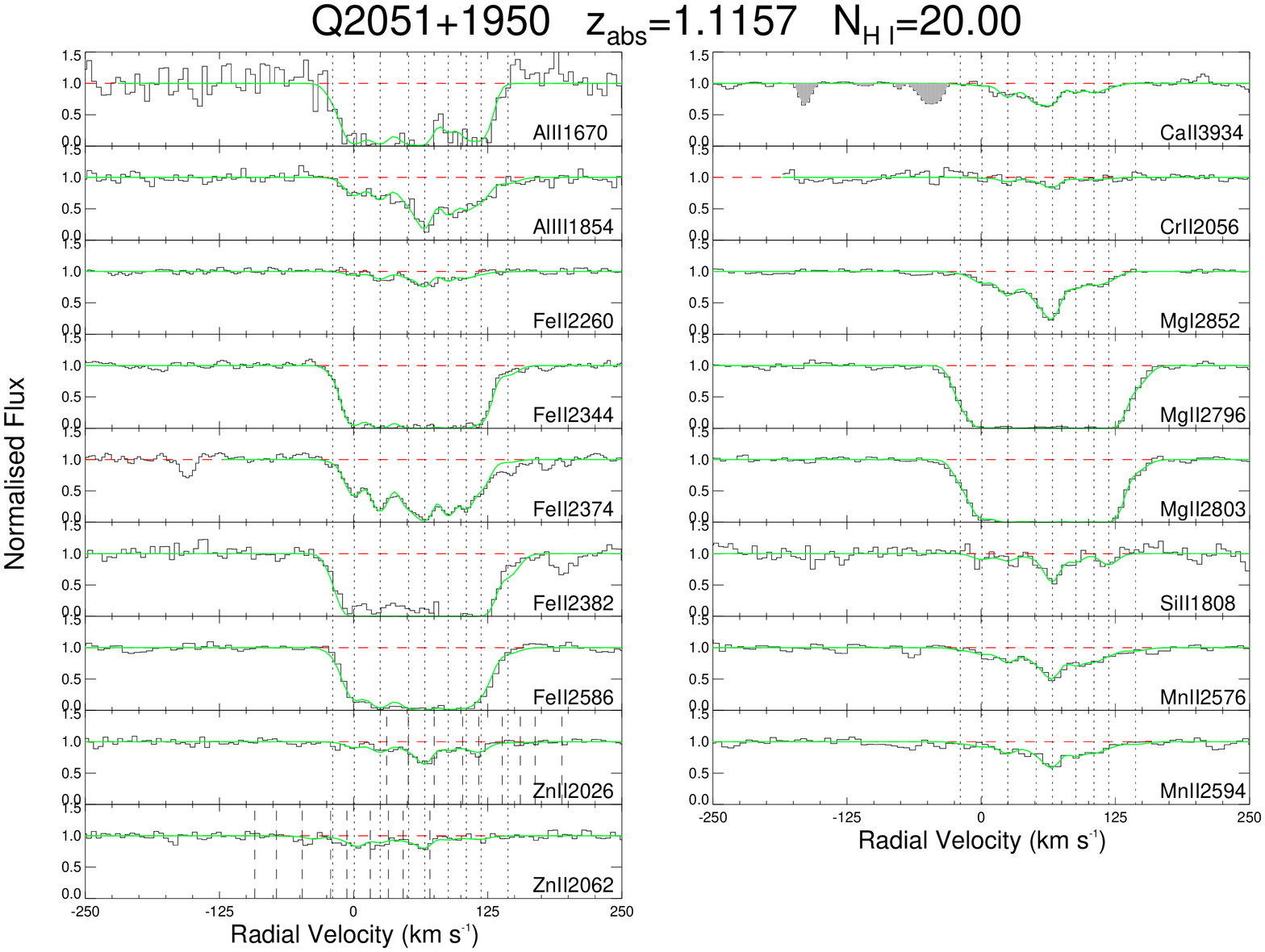}
\caption{  Same as Figure  \ref{Fig:Q0005} but for Q2051+1950  \label{Fig:Q2051}}
\includegraphics[angle=90, width=5.5in]{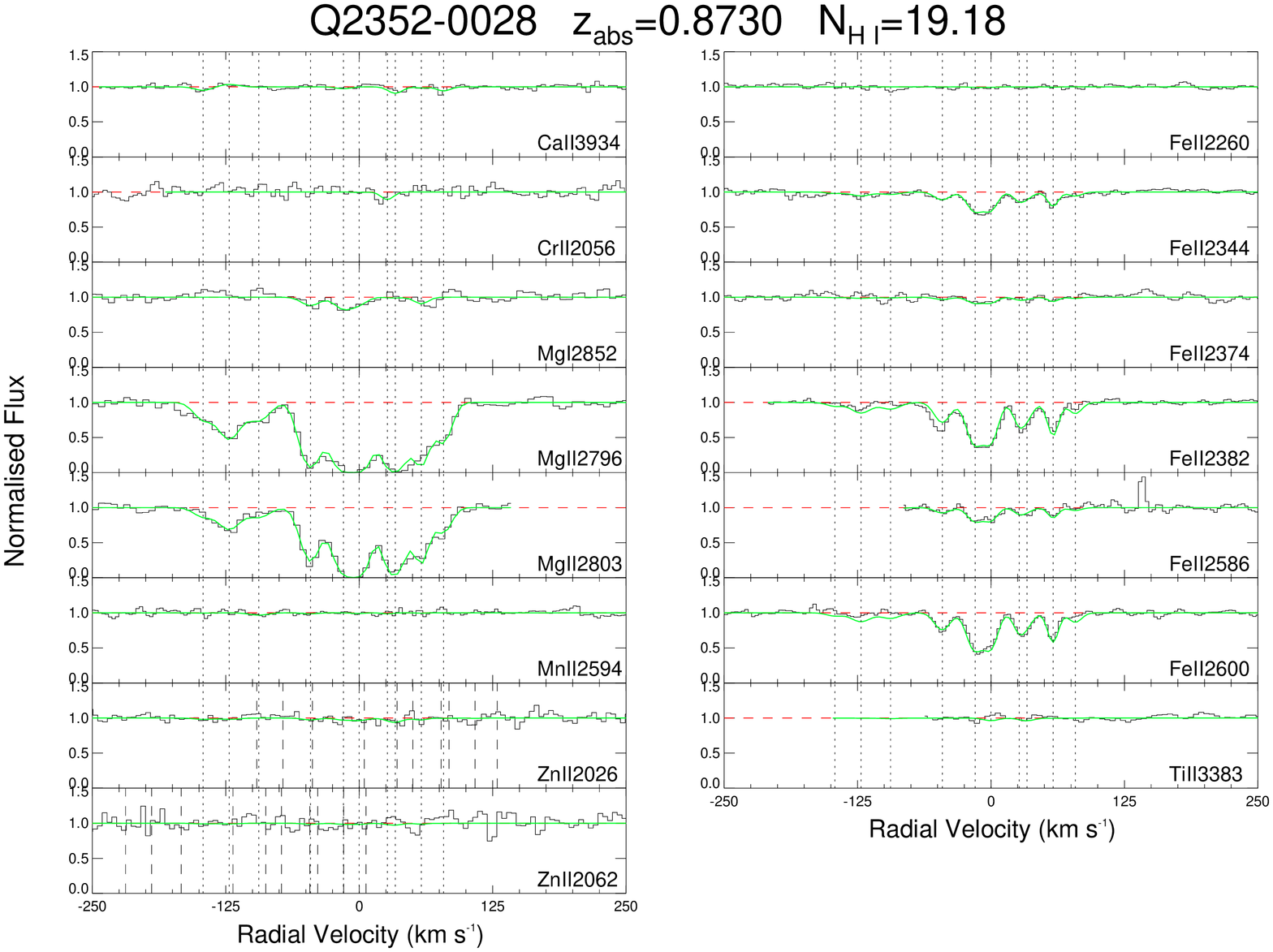}
\end{center}
\caption{ Same as Figure  \ref{Fig:Q0005} but for Q2352-0028A  \label{Fig:Q2352A} }
\end{minipage}
\end{figure*}

\begin{figure*} 
\begin{minipage}{7in}
\begin{center}
\includegraphics[angle=90, width=5.5in]{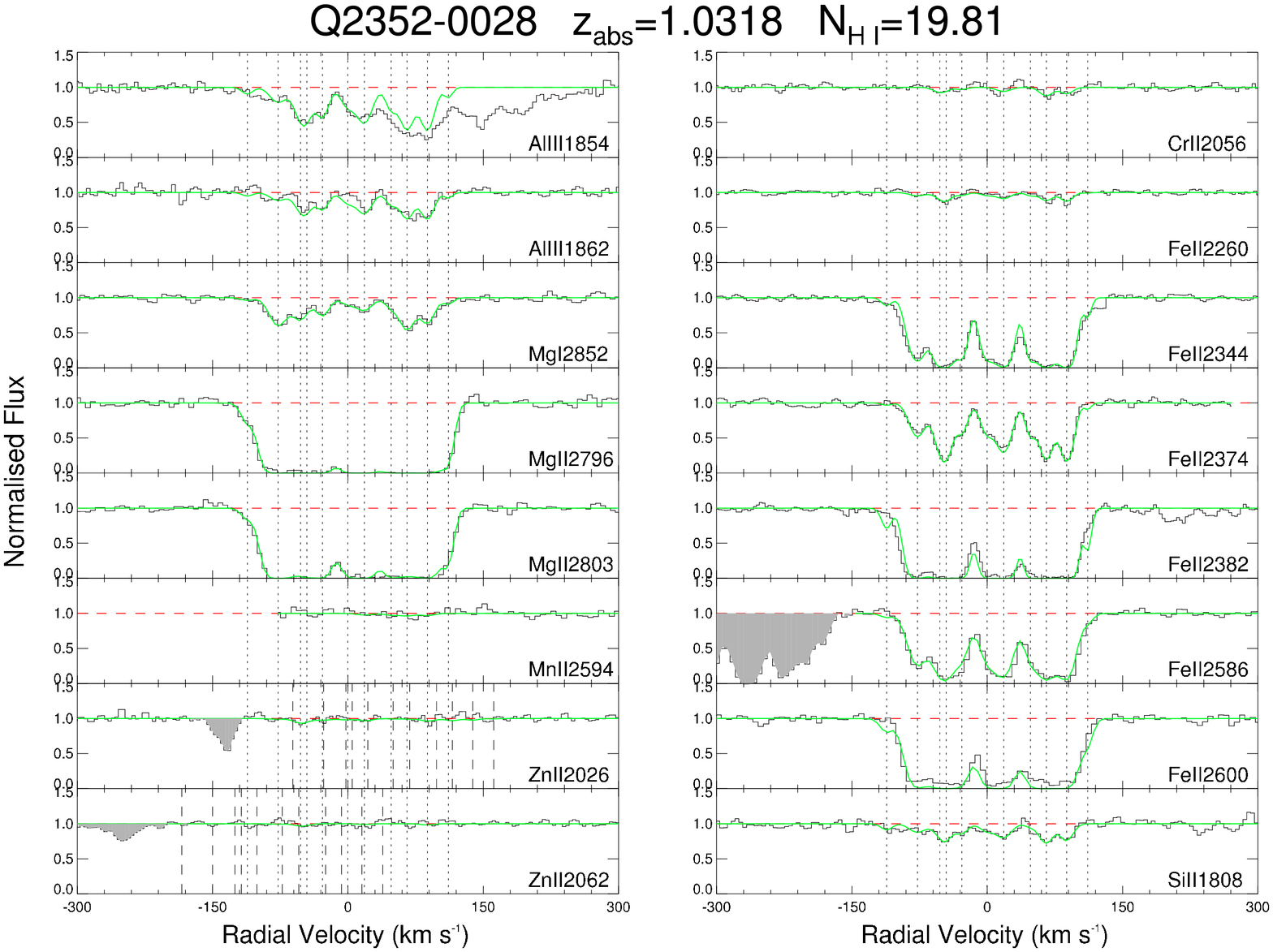}
\caption{  Same as Figure  \ref{Fig:Q0005} but for Q2352-0028B  \label{Fig:Q2352B}}
\includegraphics[angle=90, width=5.5in]{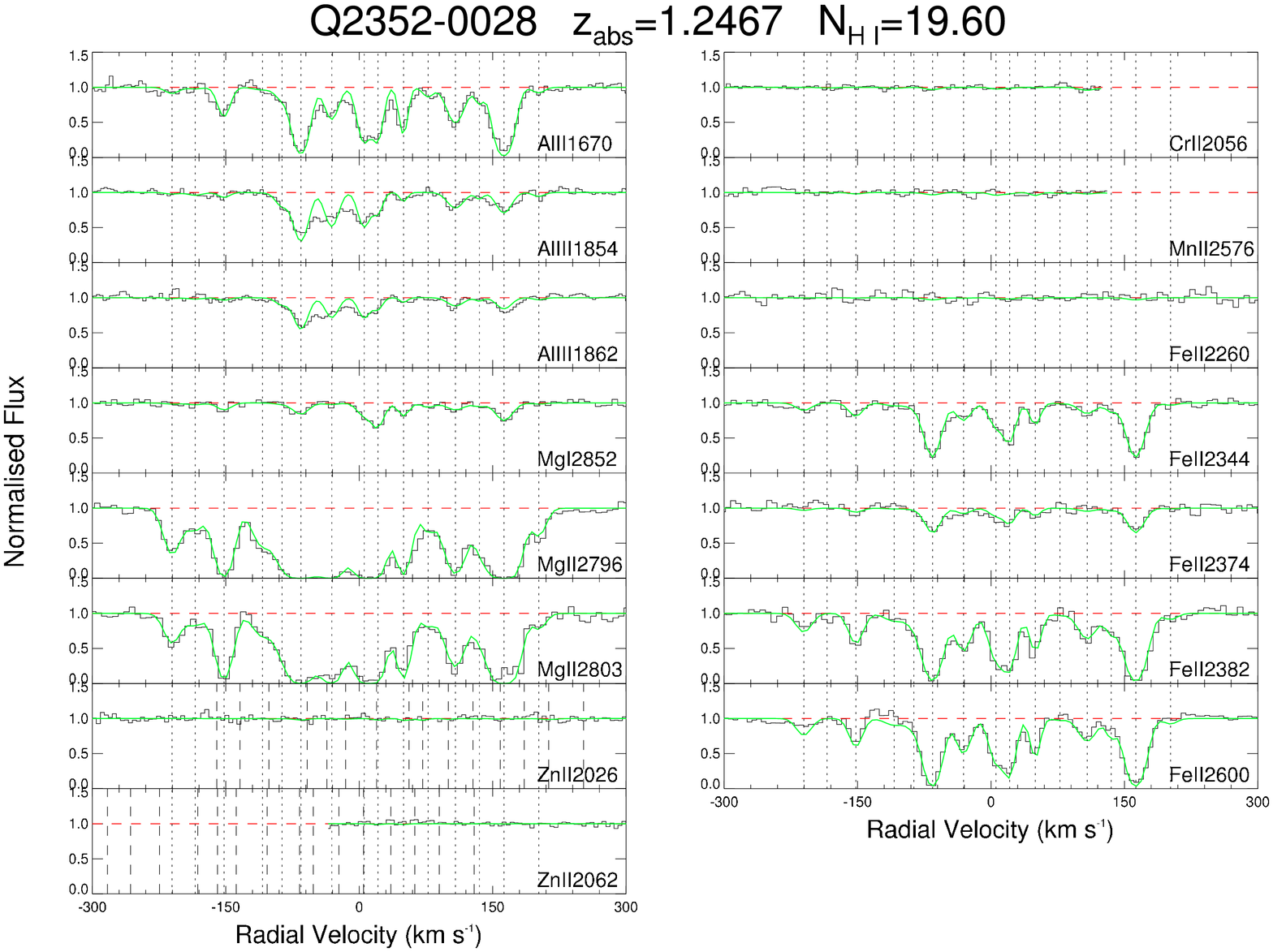}
\end{center}
\caption{ Same as Figure  \ref{Fig:Q0005} but for Q2352-0028C  \label{Fig:Q2352C} }
\end{minipage}
\end{figure*}

\subsubsection{Q0021+0104 ($z_{em}=1.829$)}
\textbf{(System A: \za=1.3259):} This is a sub-DLA system with log \nhI=20.04 \citep{Rao06}. We detect lines of Mg I $\lambda$ 2852, Mg II \lala 2796, 2803, Al II $\lambda$ 1670, Al III \lala 1854, 1862, Si II $\lambda$ 1526 and 
Fe II \lala 2344, 2374, 2382, 2586, 2600. The Mn II $\lambda$ 2576 was also possibly detected at a $\sim2 \sigma$ level. The complex absorption profile required profile required 13 components 
for an adequate fit. The Si II $\lambda$ 1808 and Al III \lala 1854, 1862 lines fell in portions of the detector with serious cosmetic issues, so they could not be measured. No Zn II lines are present
in the spectra with S/N$\sim$25 in the region. The metallicity based on Zn is thus [Zn/H]$<-$1.19. Simultaneous fits to the Fe II lines provide a measure of the Fe metallicity as 
[Fe/H]=$-0.82\pm0.11$. Due to the saturation of the Si II $\lambda$ 1526 line, only a lower limit could be placed on the column density. Even with this lower column density, the system shows signs 
of $\alpha$ enhancement with [Si/Fe]$>$+0.14. Velocity plots of several lines are shown in Figure \ref{Fig:Q0021A}. 

\noindent \textbf{(System B: \za=1.5756):} This system is a DLA with log \nhI=20.48 \citep{Rao06}. The complex velocity profile spanned $\sim$600 \kms in velocity space, and required 20 components in the fitting. 
Lines of Mg I $\lambda$ 2852, Mg II \lala 2796, 2803, Al III \lala 1854, Si II $\lambda$ 1526 and Fe II \lala 1608, 2344, 2374, 2382, 2586, 2600. The Fe II \lala 1608, 2374 lines were weak enough that 
accurate column densities could be determined. This DLA has a metallicity from Fe of [Fe/H]$=-1.34\pm0.15$, typical of other DLA values in this redshift range. No Zn II \lala 2026, 2062 lines
were detected in this system, and an upper limit of [Zn/H]$<-1.16$ was placed on the system. The Si II $\lambda$ 1526 line was detected, although saturated. We derived a Si abundance for this system of
[Si/H]=$-1.14$, which is consistent with the non-detection of the Si II $\lambda$ 1808 line. This system shows weak $\alpha$ enhancement with [Si/Fe]=+0.20, although this could also be due
to differential dust depletion. Kinematically, this system has an interesting absorption profile with two strong clusters of components separated by $\sim$100 \kms, possibly indicating a merging system. 
Velocity plots of several lines are shown in Figure \ref{Fig:Q0021B}. 

\subsubsection{Q0427$-$1302 ($z_{em}=2.166 $)} 
\textbf{(System A: \za=1.4080):} This is a weak sub-DLA system with log \nhI=19.04 \citep{Rao06}. Only three components were used in the fit, with the vast majority of the absorption in the $v\sim0$ \kms 
component. We detected lines of Mg II \lala 2796, 2803, Al II $\lambda$ 1670, Si II $\lambda$ 1526, and Fe II \lala 2344, 2374, 2382, 2586, 2600. This system is metal poor, with [Fe/H]=$-1.12\pm0.04$.
No Zn II \lala lines were detected with S/N$\sim$55 in the region, and an upper limit of [Zn/H]$<-0.58$ was placed on the system. The Al III \lala 1854, 1862 lines were covered but not detected. 
An upper limit was placed on the  Al III/Al II ratio of Al III/Al II$< - 1.14$. This system also shows slight $\alpha$ enhancement with [Si/Fe]=+0.16$\pm$0.03.
Velocity plots of several lines are shown in Figure \ref{Fig:Q0427}. 
 
\subsubsection{Q1631+1156 ($z_{em}=1.792 $)} 
\textbf{(System A: \za=0.9004):} This system is a sub-DLA with log \nhI=19.70 \citep{Rao06}. The Fe II \lala 2344, 2374 lines were affected by a cosmetic 
defect in the chip, but the Fe II \lala 2382, 2586, 2600 lines
were detected. Other lines detected were Mg I $\lambda$ 2852, Mg II \lala 2796, 2803,
and Ca II 3934. The Zn II \lala 2026, 2062 lines were covered but not detected, with S/N$\sim$6 in the region. 
Six components were used in the profile fits. We determined the metallicity for
this system based on Fe to be [Fe/H]=$-1.06\pm0.06$, and based on Zn, [Zn/H]$<-0.15$. 
Velocity plots of several lines are shown in Figure \ref{Fig:Q1631}.

\subsubsection{Q2051+1950 ($z_{em}=2.367 $)}
\textbf{(System A: \za=1.1157):} Although \citet{Rao06} determined log \nhI=19.26 for this object, we refit the UV STIS spectrum for this object to determine log \nhI=20.00$\pm$0.15. Even with this
substantially higher \nhI value, this object appears to have super solar metallicity.
We detected lines of Mg I $\lambda$ 2852, Mg II \lala 2796, 2803, Al III \lala 1854, Si II $\lambda$ 1808, Cr II $\lambda$ 2056, 
Mn \lala 2576, 2594, 2606, Fe II \lala 2260, 2344, 2374, 2382, 2586, 2600, and Zn II \lala 2026, 2062. Nine components were used in the profile fits. The Al II $\lambda$ 1670 line is heavily saturated, so only
a lower limit could be placed on the column density. The Al III/Al II ratio for this system is Al III/Al II$<-0.17$. The Ca II $\lambda$ 3934 line is clearly detected, 
with W$_0$(3934)=133$\pm$34 m\ang. 
This system shows strong Fe II absorption features, with [Fe/H]=$-0.45\pm0.15$, much higher than typical DLA systems.
We also detect Zn II \lala 2026, 2062 at $>5\sigma$, with [Zn/H]=$+0.27\pm0.18$. 
This system has moderate dust depletion with [Zn/Fe]=+0.72$\pm$0.10. Manganese is also slightly overabundant relative to Fe in this system, with [Mn/Fe]=+0.16$\pm$0.03. 
Velocity plots of several lines are shown in Figure \ref{Fig:Q2051}. 

\subsubsection{Q2352-0028 ($z_{em}=1.628 $)}
\textbf{(System A: \za=0.8739):} This is a weak sub-DLA system with log \nhI=19.18 \citep{Rao06}. 10 components were used in the profile fits. We detect lines of 
Mg I $\lambda$ 2852, Mg II \lala 2796, 2803, and Fe II \lala  2344, 2382, 2586, 2600. The Ti II $\lambda$ 3383 line was covered, but was not detected. This is not unsurprising as this line is 
weak in most systems. The Al II $\lambda$ 1670 and Al III \lala 1854, 1862 lines were below the covered wavelengths and could not be measured. The Zn II \lala 2026, 2062 lines were not detected at S/N$\sim$20
in the region. This system has a low metallicity, with [Fe/H]=$-1.17\pm0.9$, and [Zn/H]$<-0.14$.
Velocity plots of several lines are shown in Figure \ref{Fig:Q2352A}. 

\noindent \textbf{(System B: \za=1.0318):} This is a sub-DLA system with log \nhI=19.81 \citep{Rao06}. 11 components were used in the profile fitting analysis. Lines of 
Mg I $\lambda$ 2852, Mg II \lala 2796, 2803, Al III \lala 1854, 1862, Si II $\lambda$ 1808 and Fe II \lala  2260, 2344, 2374, 2382, 2586, 2600. The Al III $\lambda$ 1854 was partially blended with
an interloping feature, but the weaker Al III $\lambda$ 1862 was unperturbed. The Al II $\lambda$ 1670 line lies just below the accessible wavelengths and was not covered. This system has a moderate 
value of Fe II/Al III=+1.50$\pm$0.03. This sub-DLA has a high metallicity, with [Fe/H]=$-0.37\pm0.13$. Based  on Si the metallicity is [Si/H]=+0.14$\pm$0.14. The Zn II \lala 2026, 2062 lines 
were not detected at S/N$\sim$30 in the region, giving an upper limit on the Zn abundance as [Zn/H]$<-$0.51. This system shows signs of moderate $\alpha$ enhancement with [Si/Fe]=+0.51$\pm$0.03.
Velocity plots of several lines are shown in Figure \ref{Fig:Q2352B}. 

\noindent \textbf{(System C: \za=1.2467):} This system is a sub-DLA with log \nhI=19.60 \citep{Rao06}. The complex absorption profile required 15 components to properly fit the system. 
We detected lines of Mg I $\lambda$ 2852, Mg II \lala 2796, 2803, Al II $\lambda$ 1670, Al III \lala 1854, 1862, and Fe II \lala   2344, 2374, 2382, 2586, 2600.
This sub-DLA appears to have significant amounts of ionisation with Al III/Al II=$-0.10\pm0.03$ and Fe II/Al III=+0.82$\pm$0.03. The Zn II \lala 2026, 2062 lines were not 
detected with S/N$\sim$30 in the region. We placed an upper limit of [Zn/H]$<-0.70$ for this system. Base on the Fe II lines, this system has a metallicity of [Fe/H]=$-0.86\pm0.24$. 
Velocity plots of several lines are shown in Figure \ref{Fig:Q2352C}.

\setcounter{table}{3}
\setlength{\tabcolsep}{3pt}
\begin{sidewaystable}
\begin{minipage}{200mm}
\begin{tabular}{lccccccccccccc}
\hline
\hline	
QSO		&z$_{abs}$	&	log \nhI			&	Mg I			&	Mg II		&	Al II		&	Al III			&	Si II			&	Ca II			&	Ti II		&	Cr II		&	Mn II			&	Fe II				&	Zn II		\\
\hline
Q0005+0524	&	0.8514	&	19.08$\pm$0.04			&	12.18$\pm$0.02		&	$>$14.32	&	-		&	13.11$\pm$0.04		&	-			&	-			&	$<$11.03	&	$<$11.88	&	$<$11.01		&	13.79$\pm$0.01			&	$<$11.24	\\
AOD		&		&					&	12.24$\pm$0.04		&	$>$14.13	&			&	13.13$\pm$0.06		&				&				&			&			&				&	13.75$\pm$0.02			&			\\
Q0012$-$0122	&	1.3862	&	20.26$\pm$0.02			&	11.73$\pm$0.03		&	$>$14.09	&	$>$13.08	&	12.89$\pm$0.02		&	14.43$\pm$0.08		&	-			&	$<$11.96	&	$<$11.89	&	$<$11.41		&	14.24$\pm$0.01			&	$<$11.55	\\
AOD		&		&					&	11.75$\pm$0.05		&	$>$13.81	&	$>$13.07	&	12.83$\pm$0.04		&	14.45$\pm$0.04		&				&			&			&				&	14.25$\pm$0.01			&			\\
Q0021+0104A	&	1.3259	&	20.04$\pm$0.11			&	12.16$\pm$0.04		&	$>$14.86	&	$>$13.71	&	-			&	$>$14.90		&	-			&	 -		&	$<$12.21	&	$<$11.87		&	14.69$\pm$0.01			&	$<$11.48	\\
AOD		&		&					&	12.26$\pm$0.06		&	$>$14.50	&	$>$13.68	&				&	$>$14.86		&				&			&			&				&	14.68$\pm$0.04			&			\\
Q0021+0104B	&	1.5756	&	20.48$\pm$0.15			&	12.61$\pm$0.03		&	$>$14.57	&	-		&	12.26$\pm$0.08		&	14.88$\pm$0.03		&	-			&	  -		&	$<$12.58	&	$<$11.90		&	14.61$\pm$0.02			&	$<$11.95	\\
AOD		&		&					&	12.60$\pm$0.06		&	$>$14.51	&			&	12.31$\pm$0.17		&	14.86$\pm$0.02		&				&			&			&				&	14.58$\pm$0.02			&			\\
Q0427$-$1302	&	1.4080	&	19.04$\pm$0.04			&	-			&	$>$13.74	&	12.20$\pm$0.03	&	$<$11.06		&	13.59$\pm$0.03		&	-			&	-		&	$<$11.87	&	$<$11.65		&	13.36$\pm$0.01			&	$<$11.09	\\
AOD		&		&					&				&	$>$13.25	&	12.21$\pm$0.03	&				&	13.56$\pm$0.04		&				&			&			&				&	13.33$\pm$0.02			&			\\
Q1631+1156	&	0.9004	&	19.70$\pm$0.04			&	12.35$\pm$0.06		&	$>$14.18	&	-		&	-			&	-			&	12.13$\pm$0.07		&	$<$11.66	&	$<$12.65	&	$<$12.54		&	14.11$\pm$0.02			&	$<$12.18	\\
AOD		&		&					&	12.44$\pm$0.05		&	$>$14.04	&			&				&				&	12.22$\pm$0.10		&			&			&				&	14.17$\pm$0.03			&			\\
Q2051+1950	&	1.1157	&	20.00$\pm$0.15			&	12.64$\pm$0.02		&	$>$14.61	&	$>$13.70	&	13.53$\pm$0.03		&	15.15$\pm$0.07		&	12.55$\pm$0.04		&	-		&	12.89$\pm$0.10	&	13.21$\pm$0.02		&	15.02$\pm$0.02			&	12.90$\pm$0.10	\\
AOD		&		&					&	12.67$\pm$0.02		&	$>$14.44	&	$>$13.77	&	13.52$\pm$0.02		&	15.31$\pm$0.12		&	12.59$\pm$0.04		&			&	13.08$\pm$0.10	&	13.24$\pm$0.04		&	15.00$\pm$0.02			&	12.97$\pm$0.07	\\
Q2352$-$0028A	&	0.8730	&	19.18$\pm$0.09			&	11.85$\pm$0.06		&	$>$14.27	&	-		&	-			&	-			&	$<$11.01		&	$<$11.36	&	$<$12.37	&	$<$11.46		&	13.48$\pm$0.02			&	$<$11.67	\\
AOD		&		&					&	11.84$\pm$0.14		&	$>$14.05	&			&				&				&				&			&			&				&	13.47$\pm$0.02			&			\\
Q2352$-$0028B	&	1.0318	&	19.81$\pm$0.13			&	12.53$\pm$0.02		&	$>$14.94	&	-		&	13.41$\pm$0.02		&	15.49$\pm$0.03		&	-			&	-		&	12.96$\pm$0.06	&	$<$11.87		&	14.91$\pm$0.01			&	$<$11.93	\\
AOD		&		&					&	12.60$\pm$0.03		&	$>$14.58	&			&	13.41$\pm$0.05		&	15.49$\pm$0.06		&				&			&	12.94$\pm$0.26	&				&	14.88$\pm$0.01			&			\\
Q2352$-$0028C	&	1.2467	&	19.60$\pm$0.24			&	12.33$\pm$0.02		&	$>$15.24	&	13.49$\pm$0.02	&	13.39$\pm$0.02		&	-			&	-			&	-		&	$<$12.17	&	$<$11.60		&	14.21$\pm$0.01			&	$<$11.53	\\
AOD		&		&					&	12.32$\pm$0.07		&	$>$14.39	&	13.42$\pm$0.02	&	13.43$\pm$0.03		&				&				&			&			&				&	14.19$\pm$0.09			&			\\
\hline
\end{tabular}
\textbf{Table 3:} Total column densities from the absorbers in this sample. \label{Tab:Col2}
\end{minipage}
\end{sidewaystable}

\setcounter{table}{4}
\setlength{\tabcolsep}{3pt}
\begin{sidewaystable}
\scriptsize
%\caption{Abundances for the absorbers in this sample. \label{Tab:Abund1}}
\begin{tabular}{lccccccccccccc}
\hline
\hline	
QSO		&	$z_{abs}$&	log \nhI	&	[Zn/H]		&	[Fe/H]			&	[Fe/Zn]		&	[Si/Fe]			&	[Ca/Fe]		&		[Cr/Fe]	&	[Mn/Fe]		&	Al III / Al II	&Mg II / Mg I		&	Mg II / Al III		&	Fe II / Al III	\\
\hline
[X/Y]$_{\sun}$	&		 &			&	$-$7.37		&	$-$4.53			&	+2.84		&	+0.07			&	$-$1.13		&	$-$1.82		&	$-$1.97		&			&			&				&			\\
\hline																							
Q0005+0524	&	0.8514	&	19.08$\pm$0.04	&	$<-$0.47	&	$-$0.76$\pm$0.05	&	$>-$0.29	&		-	&	-			&	$<-$0.09	&	$<-$0.81	&	-		&	$>$2.14		&	$>$+1.21		&	0.68$\pm$0.04	\\
Q0012$-$0122	&	1.3862	&	20.26$\pm$0.02	&	$<-$1.34	&	$-$1.49$\pm$0.03	&	$>-$0.15	&	+0.12$\pm$0.08	&	-			&	$<-$0.53	&	$<-$0.86	&	$<-$0.19	&	$>$2.36		&	$>$+1.20		&	1.35$\pm$0.03	\\
Q0021+0104A	&	1.3259	&	20.04$\pm$0.11	&	$<-$1.19	&	$-$0.82$\pm$0.11	&	$>+$0.37	&	$>$+0.14	&	-			&	$<-$0.66	&	$<-$0.85	&	-		&	$>$2.70		&	-			&	-		\\
Q0021+0104B	&	1.5756	&	20.48$\pm$0.15	&	$<-$1.16	&	$-$1.34$\pm$0.15	&	$>-$0.18	&	+0.20$\pm$0.04	&	-			&	$<-$0.21	&	$<-$0.74	&	-		&	$>$1.96		&	$>$+2.31		&	2.35$\pm$0.08	\\
Q0427$-$1302	&	1.4080	&	19.04$\pm$0.04	&	$<-$0.58	&	$-$1.15$\pm$0.04	&	$>-$0.57	&	+0.16$\pm$0.03	&	-			&	$<$+0.33	&	$<$+0.26	&	$<-$1.14	&	-		&	-			&	$>$2.30		\\
Q1631+1156	&	0.9004	&	19.70$\pm$0.04	&	$<-$0.15	&	$-$1.06$\pm$0.06	&	$>-$0.91	&	-		&	$-0.85\pm$0.07		&	$<$+0.36	&	$<$+0.40	&	-		&	$>$1.83		&	-			&	-		\\
Q2051+1950	&	1.1157	&	20.00$\pm$0.15	&	$+$0.27$\pm$0.18&	$-$0.45$\pm$0.15	&   $-$0.72$\pm$0.10	&	+0.06$\pm$0.07	&	$-1.34\pm$0.05		&	$-$0.31$\pm$0.10&	+0.16$\pm$0.03	&	$<-$0.17	&	$>$1.97		&	$>$+1.08		&	1.49$\pm$0.04	\\
Q2352$-$0028A	&	0.8730	&	19.18$\pm$0.09	&	$<-$0.14	&	$-$1.17$\pm$0.09	&	$>-$1.03	&	-		&	$<-1.34$		&	$<$+0.71	&	$<-$0.05	&	-		&	$>$2.42		&	-			&	-		\\
Q2352$-$0028B	&	1.0318	&	19.81$\pm$0.13	&	$<-$0.51	&	$-$0.37$\pm$0.13	&	$>+$0.14	&	+0.51$\pm$0.03	&	-			&	$-$0.13$\pm$0.06&	$<-$1.07	&	-		&	$>$2.41		&	$>$+1.53		&	1.50$\pm$0.03	\\
Q2352$-$0028C	&	1.2467	&	19.60$\pm$0.24	&	$<-$0.70	&	$-$0.86$\pm$0.24	&	$>-$0.16	&	-		&				&	$<-$0.22	&	$<-$0.64	&	$-0.10\pm0.03$	&	$>$2.91		&	$>$+1.85		&	0.82$\pm$0.03	\\
\hline
\end{tabular}
\begin{minipage}{200mm}
\textbf{Table 4:} Abundances and adjacent ion ratios for the absorbers in this sample. \label{Tab:Abund}
\end{minipage}
\end{sidewaystable}

%%%%%%%%%%%%%%%%%%%%%%%%%%%%%%%%%%%%%%%%%%%%%%%%%%%%%%%%%%%%%%%%%%%%%%%%%%%%%%%%%%%%%%%%%%%%%%%%
%%%%%%%%%%%%%%%%%%%%%%%%%%%%%%%%%%%%%%%%%%%%%%%%%%%%%%%%%%%%%%%%%%%%%%%%%%%%%%%%%%%%%%%%%%%%%%%%

\section{[Mn/Fe] - Nucleosynthetic Effects } \label{Sec:MnFe}
   Mn and Fe are an interesting pair of elements to study in QSO absorbers for reasons discussed below,
and has been investigated in the past by several groups \citep{Pet00, Des02, Led02}. The lines of Mn II \lala 2576, 2594, 2606 are 
usually simultaneously accessible with Fe II lines with the MIKE spectrograph due to the large wavelength coverage available. 
In Milky Way stars, the Mn abundance seems to have a strong metallicity dependence. 
A clear trend between [Mn/Fe] and [Fe/H] in the sense that [Mn/Fe] increases to near solar values as [Fe/H] increases also is seen \citep{McW03, Nis00, Grat04}. 
This seems to indicate that the nucleosynthetic 
origin of Mn is in Type Ia supernovae, as with the $\alpha$-capture elements which are produced in Type II SNe, [$\alpha$/Fe] tends to decrease with increasing [Fe/H].
\citet{Sam98} argue that $\sim75\%$ of Mn is produced in Type 1a supernovae explosions. 
On the other hand, \citet{Tim95} suggest that metallicity 
dependent yields in Type II supernovae can also reproduce the observed trends. \citet{Felt07} favor metallicity dependent 
yields with Type II SNe as the major contributors based on their observations of stellar abundances in the Milky Way and a compilation
of other stellar abundances, but note that
the Type Ia enrichment scenario is impossible to rule out. 

Whatever the underlying production mechanism of Mn may be, 
there does seem to be a trend between 
[Mn/Fe] and metallicity. As the condensation temperatures of Mn and Fe are similar, 
and accordingly have similar levels of depletion especially in the warm and halo ISM of the Milky Way, the relative abundances of these
two elements are expected to be mainly nucleosynthetic in origin.

\begin{figure}
\begin{center}
\vspace{0.25cm}
\includegraphics[angle=0, width=\linewidth]{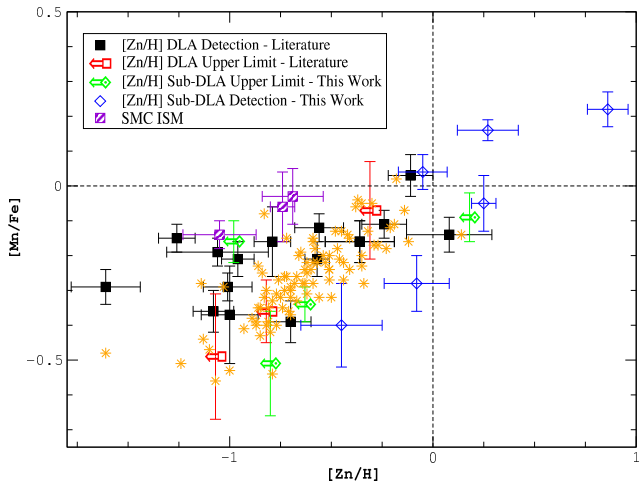}
\caption{ [Mn/Fe] vs. [Zn/H] for sub-DLAs, as well as the DLAs with Mn and Fe detections from the literature. The  star shaped orange points
are Milky Way stellar abundances from \citet{Red06}. The sub-DLA points are from these measurements as well as \citet{Mei07, Mei08}. Also shown are the ISM 
abundances for the SMC from \citet{Wel01}  \label{Fig:MnFe}}
\end{center}
\end{figure}

	In Figure \ref{Fig:MnFe} we show [Mn/Fe] vs. [Zn/H] for these absorbers, as well as ones taken from the literature. Overlayed on the same graph are the data points
from \citet{Red06} from Milky Way stars, and the interstellar abundances from \citet{Wel01}. A clear trend of increasing [Mn/Fe] with increasing [Zn/H] is seen, as is seen in the Milky Way. 
A Spearman rank correlation coefficient for these data was determined to be r$_{\rm S}$ = $0.497$ with a probability of obtaining this value by chance of 0.003. 
Kendall's $\tau$ was also determined to be $\tau=0.359$, with a probability of no correlation also of 0.003. 
The absorber sample has a larger dispersion than the stellar abundances from \citet{Red06}, which is possibly a result of the combined effects 
of some differential depletion between Mn and Fe, and the fact that the galaxies sampled via absorption lines are likely from a mixture of morphological types.
The two sub-DLA points in Figure \ref{Fig:MnFe} that lie below the stellar points at ([Zn/H], [Mn/Fe]) $\sim$($-0.5 -0.4$) and ($-0.1, -0.3$) also have the largest 
associated errors in [Mn/Fe], but are well within $\sim2\sigma$ of the stellar abundances. The apparent optical depth and profile fitting column density determinations of Zn II in these
systems agree within the error bars. 

We note that the relative abundance ratio of [Mn/Fe] does not appear to be significantly altered by ionisation effects 
for reasonable estimates of the ionisation parameter \citep{Des02, Mei08}. 
A similar trend was seen in \citet{Led02} for [Mn/Fe] as well, albeit with a smaller sample size.

%%%%%%%%%%%%%%%%%%%%%%%%%%%%%%%%%%%%%%%%%%%%%%%%%%%%%%%%%%%%%%%%%%%%%%%%%%%%%%%%%%%%%%%%%%%%%%%%
%%%%%%%%%%%%%%%%%%%%%%%%%%%%%%%%%%%%%%%%%%%%%%%%%%%%%%%%%%%%%%%%%%%%%%%%%%%%%%%%%%%%%%%%%%%%%%%%

\section{Discussion } \label{Sec:Abund}

Total ionic column densities are given in Table 3. 
The abundances for the observed systems are given in Table 4, where 
we have used the total column densities (i.e. the sum of the column densities in the individual components of a system that were determined via profile fitting method)
along with the total \nhI as given in Table \ref{Tab1}, to determine the abundances of these systems. 
We have not assumed any ionisation corrections on these abundances, and have assumed the first ions to be the dominant 
ionisation species of the elements for which these abundances have been determined, 
namely Zn, Fe, Mn, Cr, and Si. Solar systems abundances from \citet{Lodd03} are also given in Table 4.

Relative abundances of various elements are also given in Table 4, with the column densities also determined from the profile fitting analysis. Along with the metallicities, we give 
the ratio [Zn/Fe], which is often used as an indicator of dust depletion. We also provide the ratios of [Si/Fe], [Ca/Fe], [Cr/Fe], and [Mn/Fe]. 
Finally, we provide ratios of the column densities of the adjacent ions Al III/Al II and Mg II/Mg I as well as Mg II/Al III and Fe II/Al III,
any of which may provide information about the ionisation in these
systems. Photoionization modeling has shown that the ionisation corrections necessary are small in most cases, with correction terms comparable to the
individual errors on the abundances \citep{Des03, Pro06, Mei07, Mei08}.

The trend of a rising ratio in [Mn/Fe] with increasing metallicity for QSO absorbers, both DLAs and sub-DLAs, 
mimics the trend seen in Milky Way stars. A similar comparison with stellar abundances
from the SMC and LMC could also shed light on the morphological types of the absorber host galaxies, as the Magellanic clouds have undergone 
much different star formation histories than the Milky Way \citep{PT98, Carr08}. 

Similarly, the [Mn/Fe] ratio would be interesting to study in detail for a large number of systems 
over a large redshift range as was done for small samples in \citet{Des02, Led02}. Due to the intrinsic scatter in the data, a large statistical sample 
is needed to  investigate [Mn/Fe] vs. $z$.
The Mn II triplet at $\sim$2600 \ang is however
difficult to detect at $z>1.75$ as these lines are sometimes blended with 
telluric features at these redshifts. On the other hand, these lines can be probed with ground based spectra at $z \ga 0.25$. 

It would be interesting to see of the [Mn/Fe] ratio plateaus to a constant value of
[Mn/Fe]$\sim-0.4$ dex as seen in metal poor stars with [Fe/H]$\la-$1.5 \citep{Bai04, McW03}.
Also, [Mn/O] would be an interesting abundance ratio to study as O is almost solely produced in Type II SNe, and Fe is produced in 
both Type Ia and Type II explosions. [Mn/O] vs [O/H] for dwarf spheroidals and the Milky Way show quite distinct 
trends (see for instance \citealt{Felt07}), and as such may shed light on the morphology of these absorption systems. 
With these elements however, the task of separating the differential dust depletion between Mn and O
and true nucleosynthetic differences may be difficult.

Here, we have presented rest frame equivalent widths, column densities, and abundances for these 10 absorbers based on spectra taken with the MIKE spectrograph.
Zn, the preferred metallicity indicator, is detected in only one system (the \za=1.1157 system in Q2051+1950), 
so an estimate of the mean metallicity based on Zn from survival analysis 
is impossible as survival analysis requires a higher fraction of detections for an accurate estimate of the mean.

Based on the more heavily depleted element Fe which is detected in all systems,
the \nhI-weighted mean metallicity is $\langle$[Fe/H]$\rangle=-0.77\pm0.11$. This is nearly
0.7 dex higher than what is seen in DLA systems at these redshifts, $\langle$[Fe/H]$_{DLA}\rangle \sim -1.5$, and even higher than the \nhI-weighted mean metallicity 
based on Zn measurements for DLAs at these redshifts (see for instance \citet{Kul07} and references therein). 
In an forthcoming paper, we combine these values with those from our previous work in \citet{Mei07, Mei08, Per06b} 
to examine the full sample of high resolution sub-DLAs at $z < 1.5$ from our MIKE and UVES observations including kinematics,  
mean metallicities, and relative abundances.

\section*{Acknowledgments}
We thank the exceptionally helpful staff of Las Campanas Observatory for their assistance during the observing runs. Thanks to the 
anonymous referee for several helpful suggestions. 
J. Meiring and V.P. Kulkarni gratefully acknowledge support from the National Science 
Foundation grant AST-0607739 (PI Kulkarni). J. Meiring acknowledges partial support from a South Carolina Space Grant graduate student fellowship for a 
portion of this work.

\bsp

\section{Appendix}

\subsection{\nhI Determinations}

   The systems studied in this work all have known N$_{\rm H \ I}$ from HST spectra. For completeness, we provide plots of the Voigt profiles of the
Lyman-$\alpha$ transition using the best fit values of the column density from \citet{Rao06}. Due to the low resolution and S/N of
these UV spectra, only one component was used in the fits.  We show in the following figures the Voigt profiles corresponding 
to the column densities given by \citet{Rao06} and convolved with a Gaussian instrumental spread function based on a two pixel resolution element, superimposed on the
archival data from HST cycle 6 program 6577, cycle 9 program 8569, and cycle 11 program 9382. 
We note that the normalization, i.e., the continuum fit that we define may differ from that adopted by \citet{Rao06}. 
For Q2051+1950, we have revised the fit of \citet{Rao06}  to log \nhI=20.00$\pm$0.15. For all other cases, we find their values completely in agreement. 
For our continuum fits, a polynomial 
typically of order 5 or less or a cubic spline was used, and the absorption line itself was excluded from the fitting region.
Also over-plotted are profiles with H I column densities smaller and larger by 0.15 dex than the best fit values.
A bar located below the continuum level in the absorption feature denotes the range of absorption seen in the Mg II profiles in velocity space. 
For Q0005+0524, a shift of $\sim$200 \kms was applied to align the profile due to apparent inaccuracies in the wavelength calibration of the FOS spectrum. 
See also \citet{Mill99} for more on the FOS pipeline wavelength calibration inaccuracies.   

Higher resolution UV spectra covering higher Lyman series transitions (Lyman-$\beta$, Lyman-$\gamma$, etc) would help to resolve the effects of the wide spread of the Mg II components, 
and discern abundance differences between components which is not possible with the current lower resolution UV spectra.

\begin{figure*}
\begin{minipage}{152mm}
\begin{center}
$\begin{array}{c@{\hspace{0.0cm}}c}
\multicolumn{1}{l}{\mbox{\bf }} &
	\multicolumn{1}{l}{\mbox{\bf }} \\ 
		\includegraphics[angle=90,width=3in,height=2.5in]{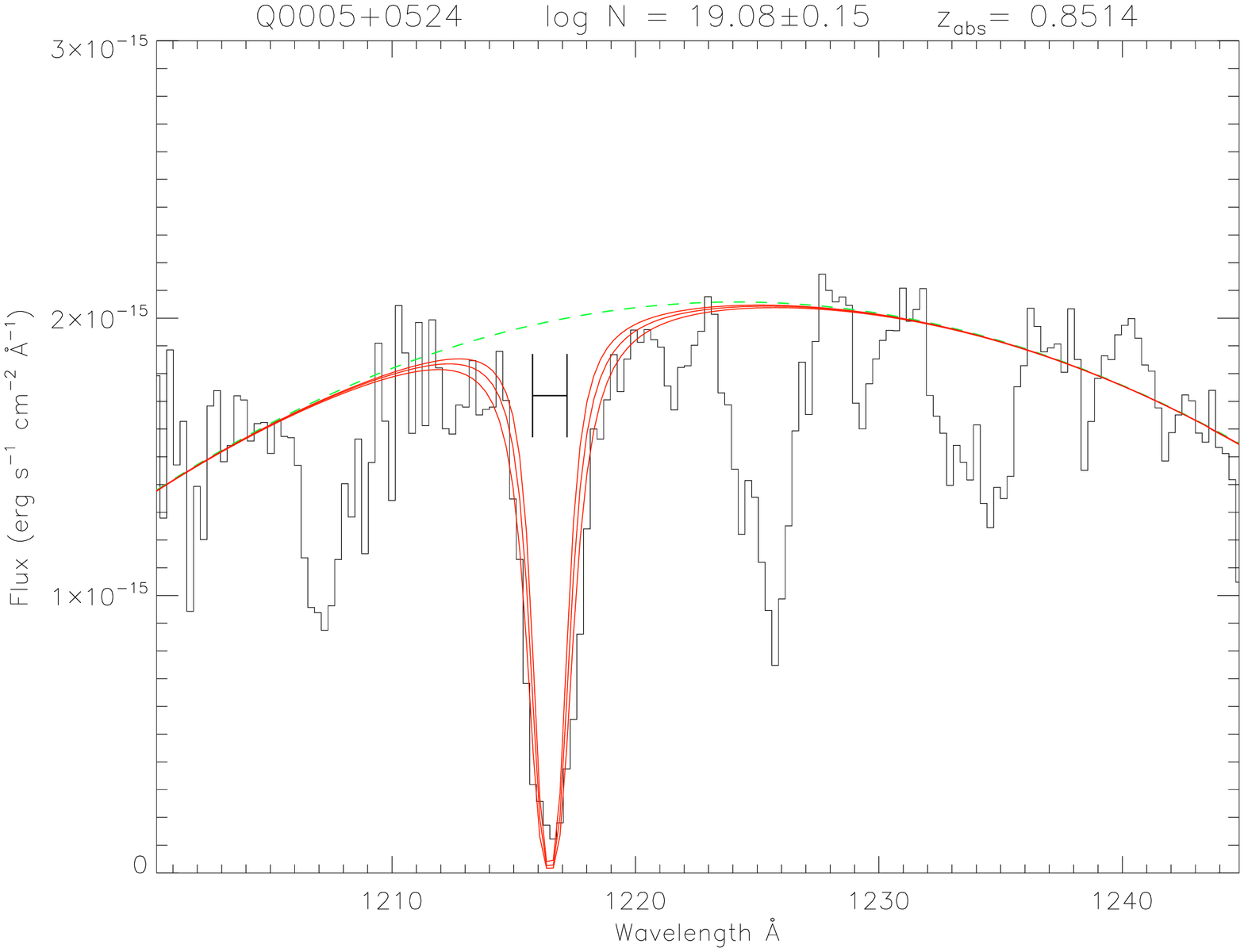} & \includegraphics[angle=90,width=3in,height=2.5in]{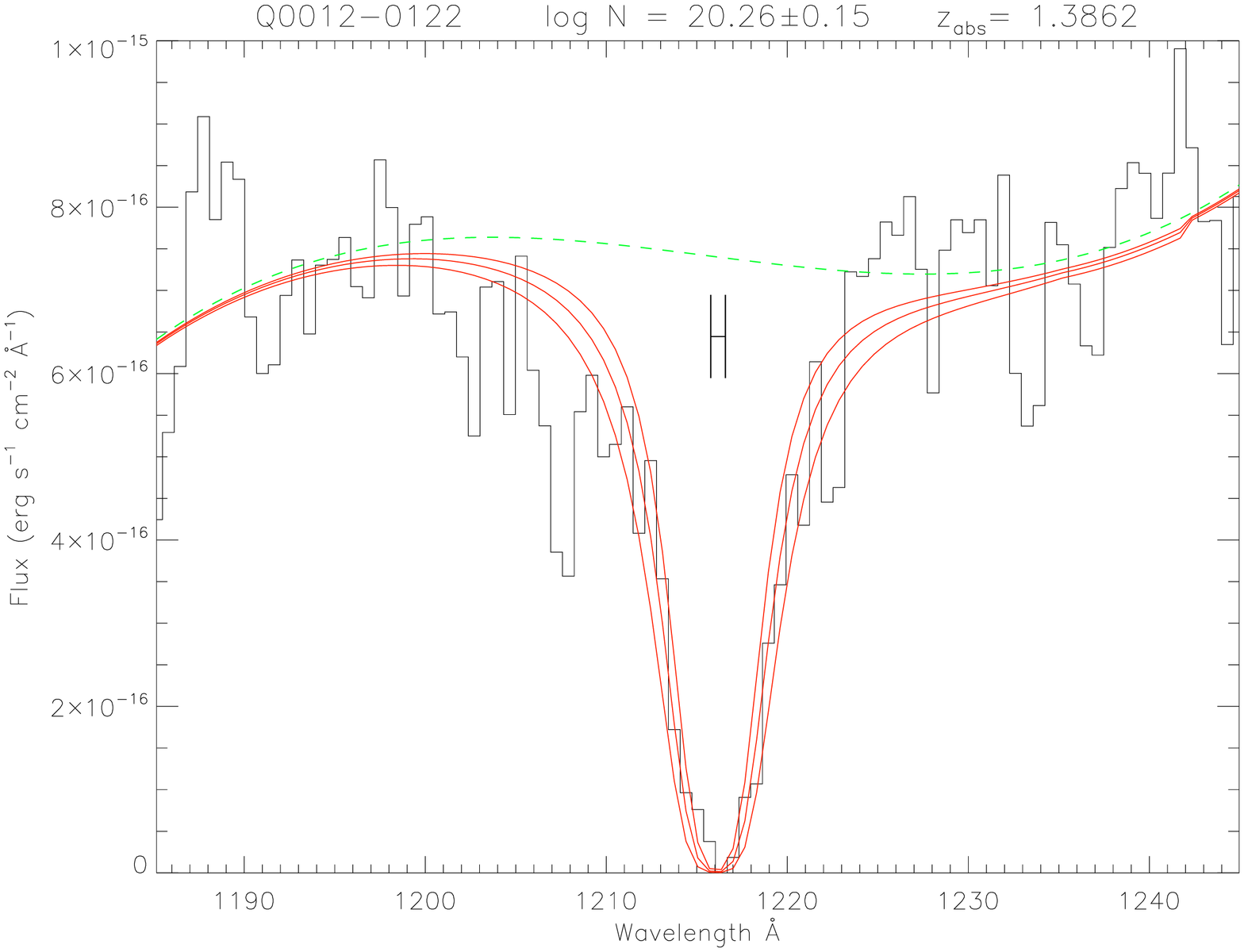} \\ [-0.0cm]
		\includegraphics[angle=90,width=3in,height=2.5in]{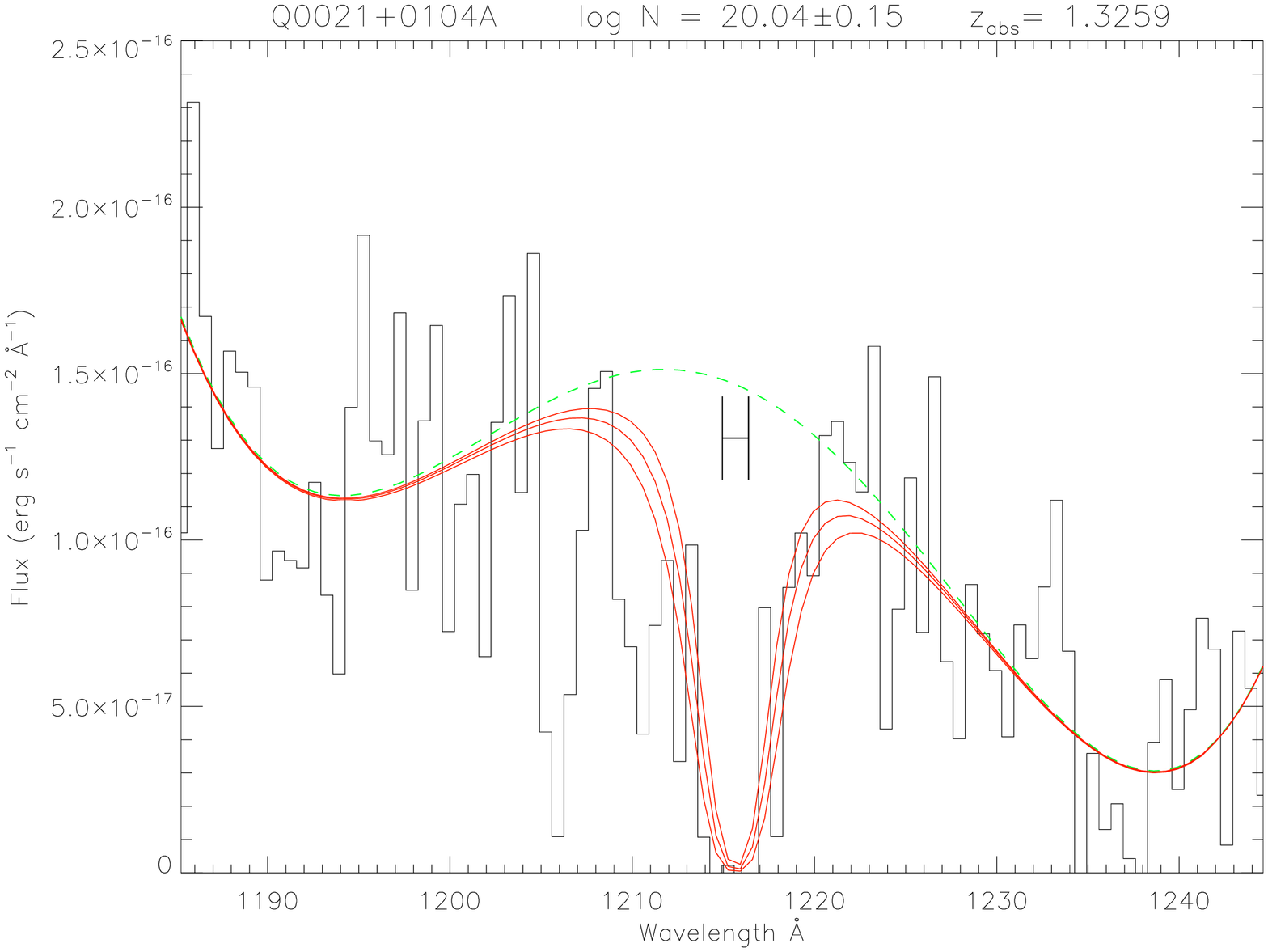} & \includegraphics[angle=90,width=3in,height=2.5in]{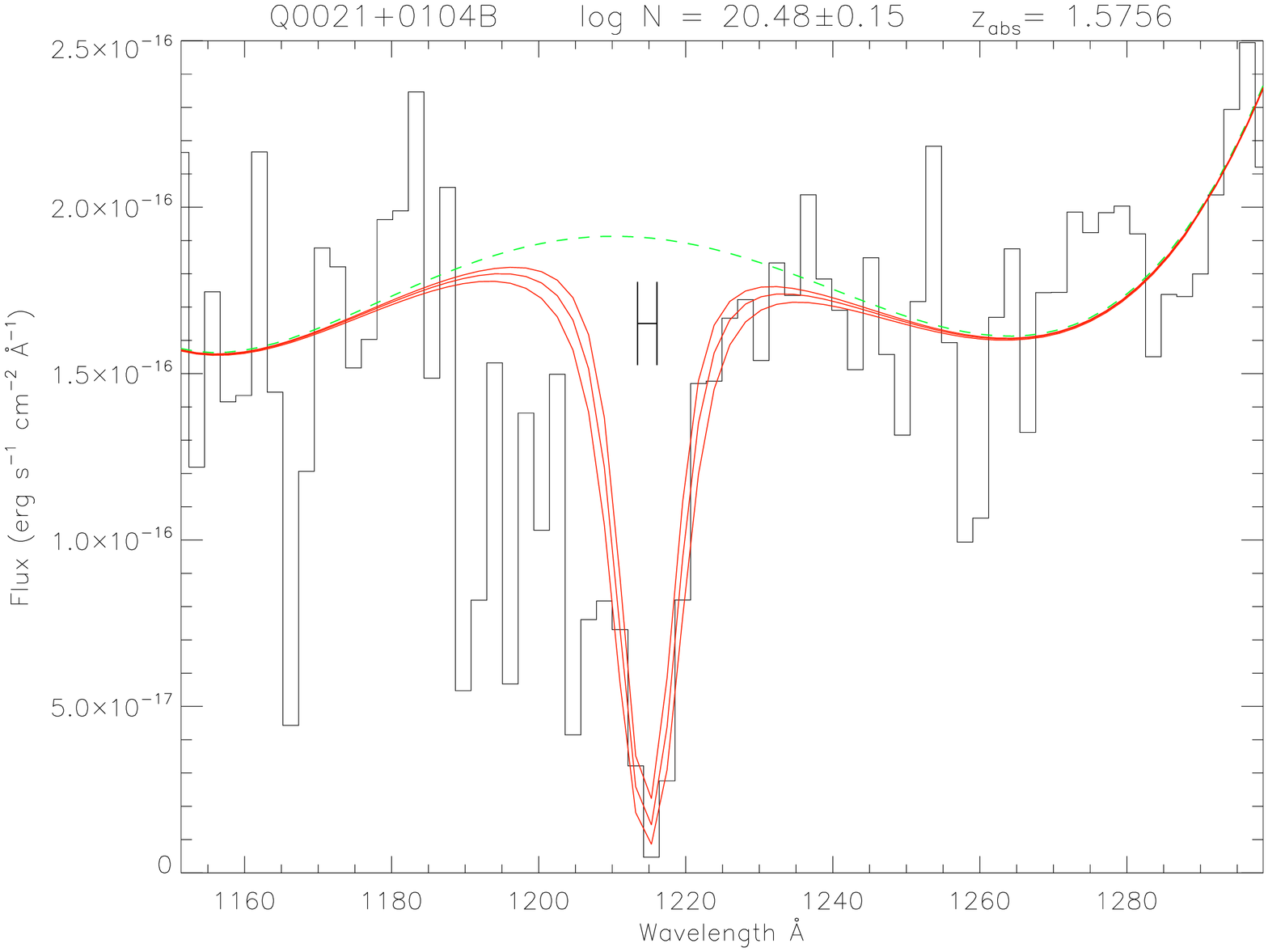} \\  [-0.0cm]
		\includegraphics[angle=90,width=3in,height=2.5in]{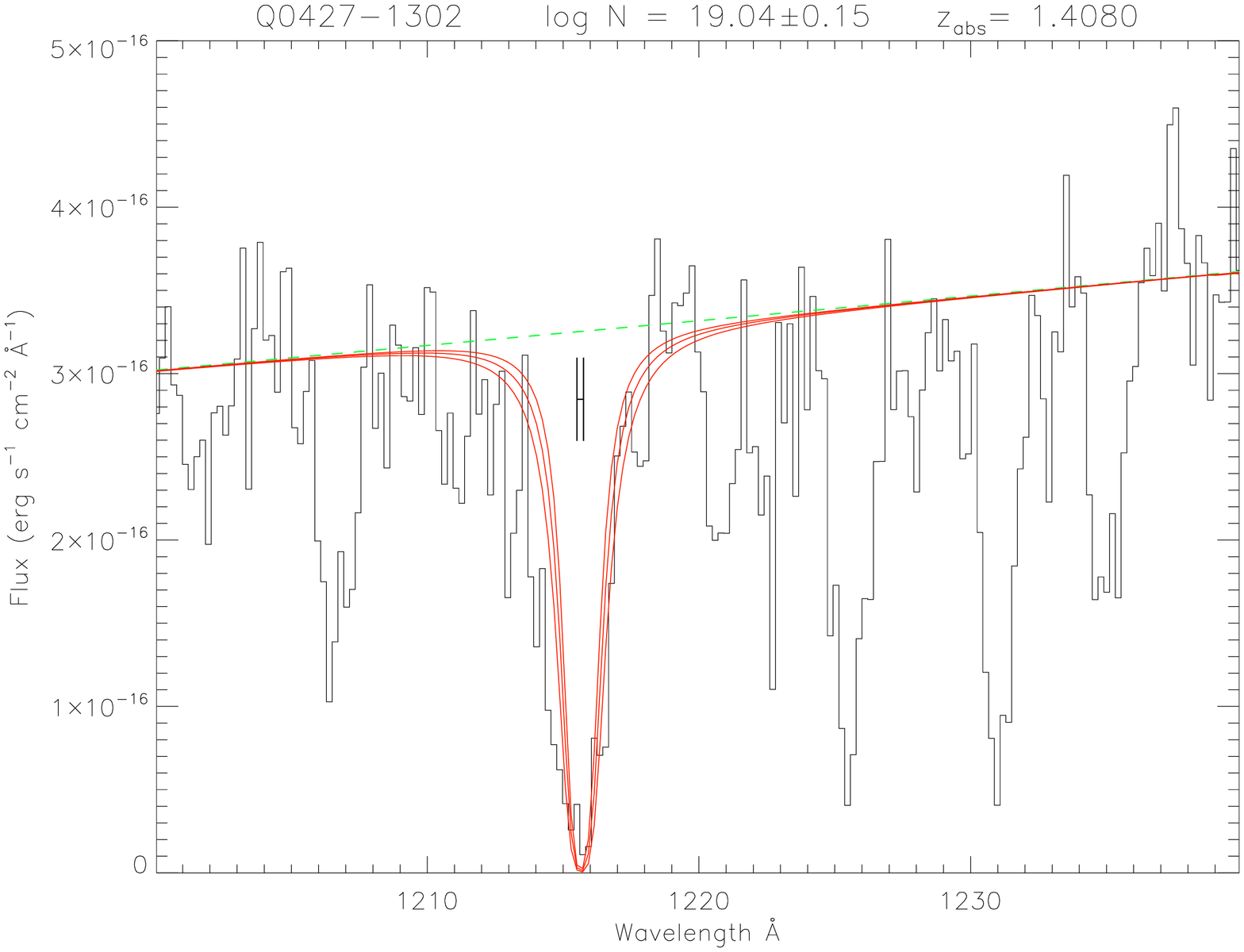} & \includegraphics[angle=90,width=3in,height=2.5in]{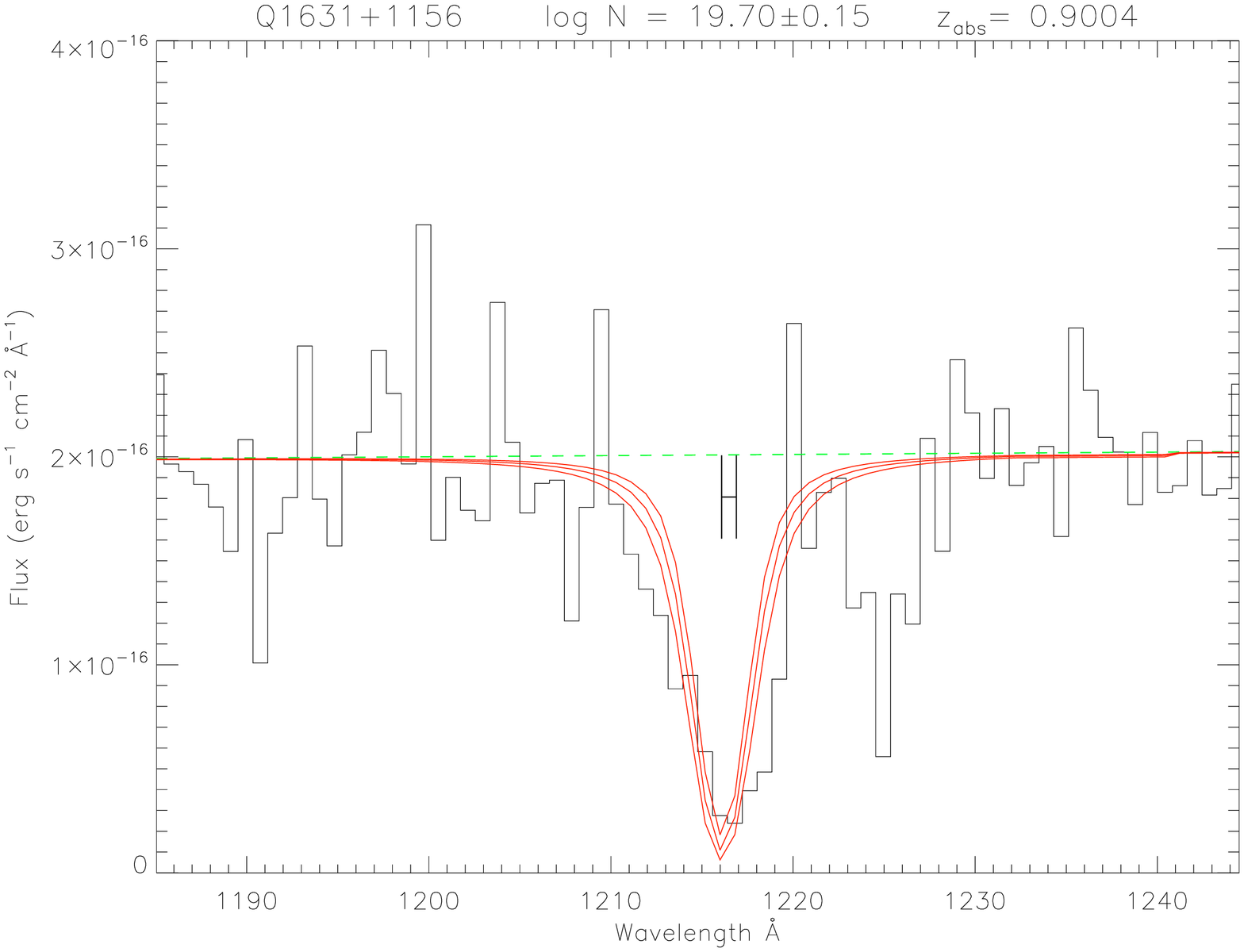} \\  [-0.0cm]
\end{array}$
\end{center}
\caption{UV spectra of the systems in this sample. 
The dashed green line indicates the best fit level of the continuum. Superimposed on the data are theoretical Voigt profiles that have been convolved with 
a gaussian instrumental spread function. The middle profile is the best fit value from
\citet{Rao06}, and the upper and lower column densities have been modified by $\pm$0.15 dex from the best fit value. The bar located below the continuum level in the 
absorption feature denotes the range of absorption seen in the Mg II profiles in velocity space. \label{Fig:UV1}}
\end{minipage}
\end{figure*}

\begin{figure*}
\begin{minipage}{152mm}
\begin{center}
$\begin{array}{c@{\hspace{0.0cm}}c}
\multicolumn{1}{l}{\mbox{\bf }} &
	\multicolumn{1}{l}{\mbox{\bf }} \\ 
		\includegraphics[angle=90,width=3in,height=2.5in]{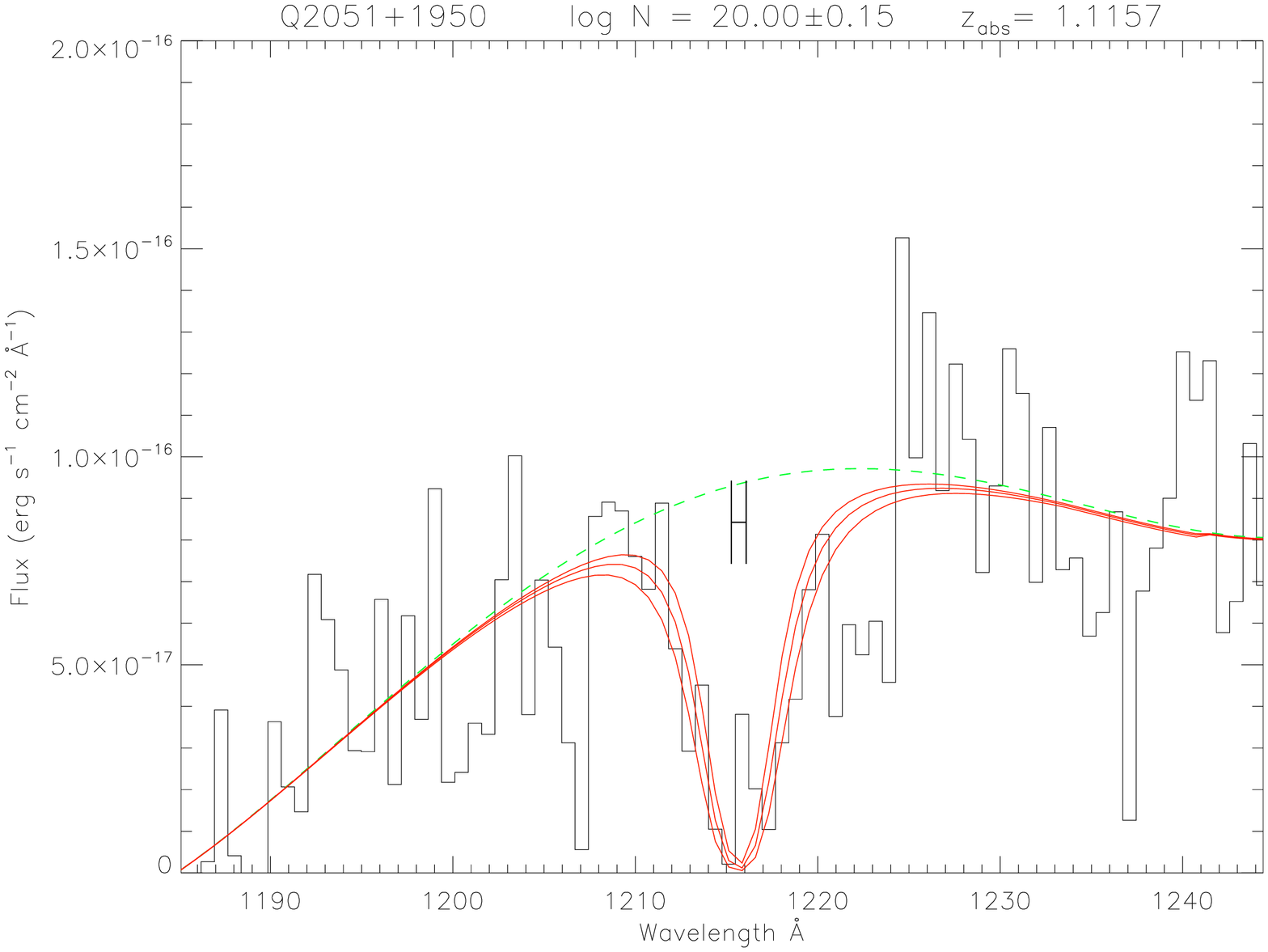} & \includegraphics[angle=90,width=3in,height=2.5in]{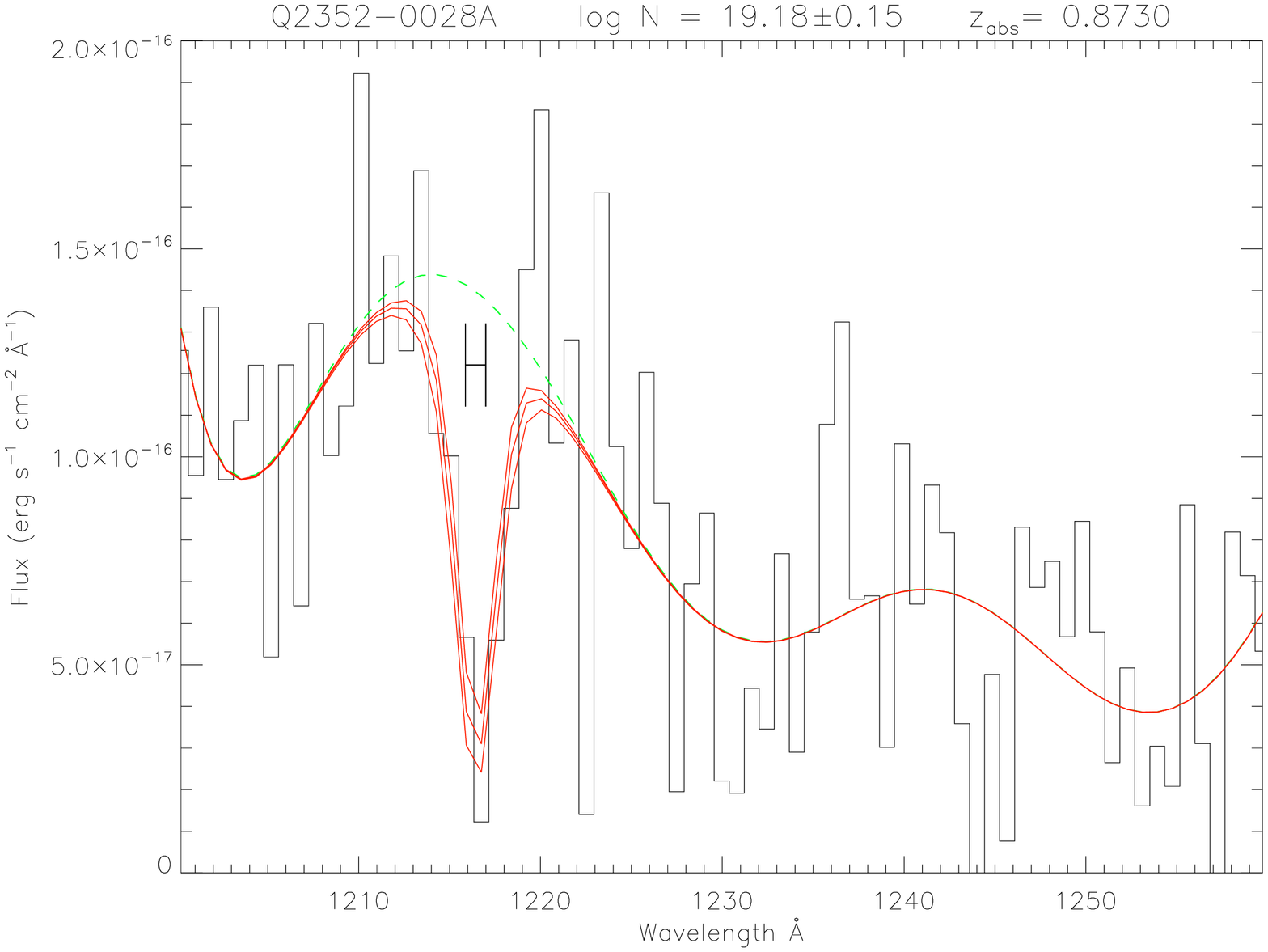} \\ [-0.0cm]
		\includegraphics[angle=90,width=3in,height=2.5in]{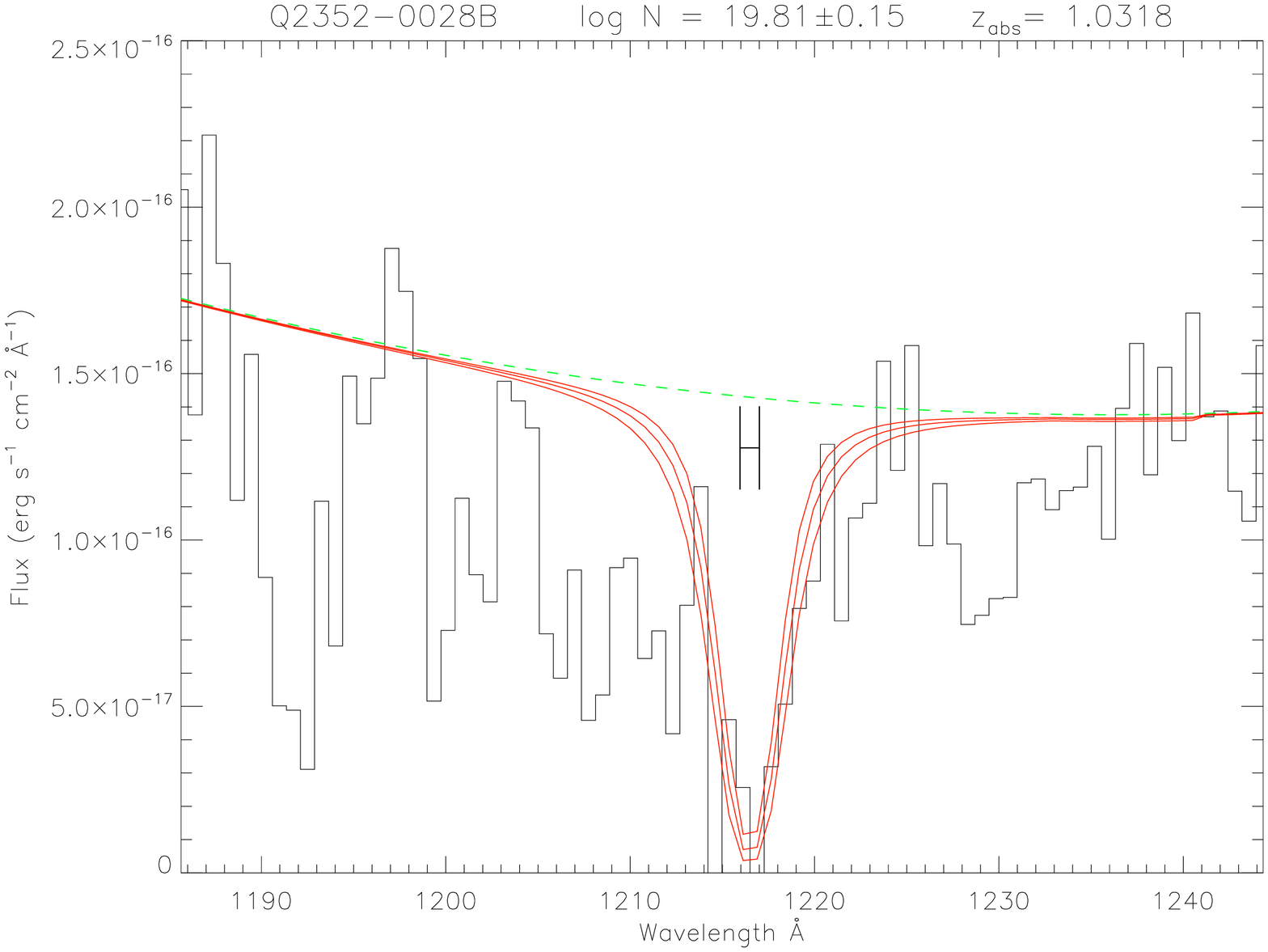} & \includegraphics[angle=90,width=3in,height=2.5in]{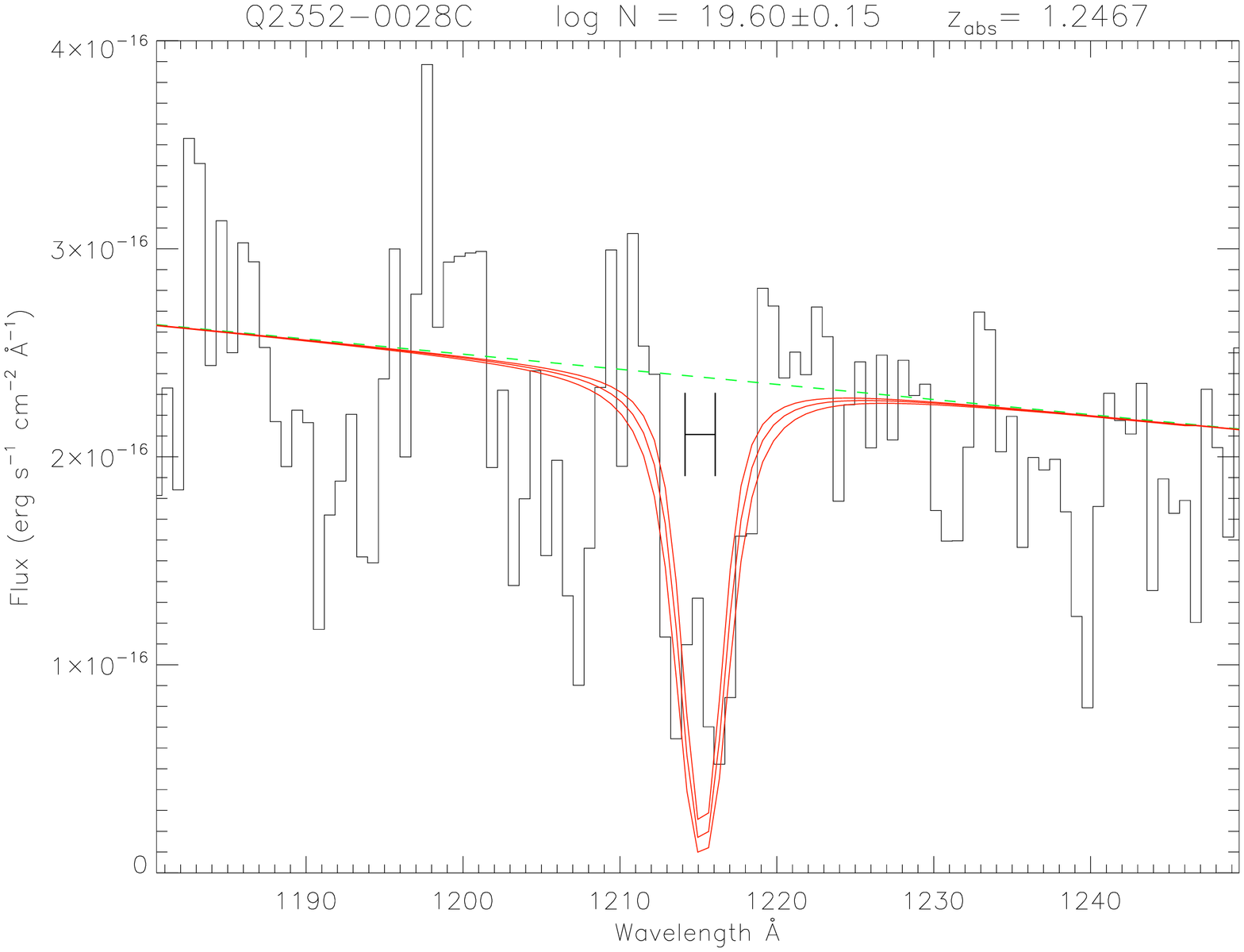} \\  [-0.0cm]
\end{array}$
\end{center}
\caption{ Same as Figure \ref{Fig:UV1} \label{Fig:UV2}}
\end{minipage}
\end{figure*}

\subsection{Individual Fit Parameters} 
Here we give the parameters for the fits for each individual system. Radial velocities and effective Doppler parameters of the components are in units of \kms, while the column densities are in cm$^{-2}$. The errors 
for each component are 1$\sigma$ formal errors from FITS6P. 

\begin{table*}
\begin{minipage}{100mm}
\caption{Column densities for Q0005+0524, z$_{abs}$=0.8514, \nhI=19.08 \label{Tab:Q0005}}
\begin{tabular}{cccccc}
\hline\hline
v	&b$_{eff}$	&	Mg I			&	MgII 			&	Al III 			&	Fe II			\\
\hline
$-$26	&	8.6	&	(4.24$\pm$0.27)E11	&	$>$3.09E13		&	(2.71$\pm$0.53)E12	&	(1.63$\pm$0.04)E13	\\
$-$11	&	6.2	&	(5.55$\pm$0.32)E11	&	$>$4.18E13		&	(4.30$\pm$0.68)E12	&	(2.87$\pm$0.07)E13	\\
7	&	5.6	&	(4.26$\pm$0.26)E11	&	$>$7.69E13		&	(1.67$\pm$0.43)E12	&	(1.02$\pm$0.03)E13	\\
35	&	10.3	&	(9.78$\pm$2.09)E10	&	(3.01$\pm$0.21)E12	&	-			&	-			\\
145	&	4.7	&	-			&	$>$8.09E12		&	(6.10$\pm$3.43)E11	&	(2.36$\pm$0.21)E12	\\
182	&	6.9	&	-			&	$>$4.72E13		&	(3.44$\pm$0.55)E12	&	(4.01$\pm$0.24)E12	\\
301	&	9.0	&	-			&	(9.11$\pm$1.39)E11	&	-			&	-			\\
412	&	5.5	&	-			&	(1.34$\pm$1.55)E12	&	-			&	-			\\
\hline
\end{tabular}
\end{minipage}
\end{table*}

\begin{table*}
\begin{minipage}{150mm}
\footnotesize
\caption{Column densities for Q0012$-$0122, z$_{abs}$=1.3862, \nhI=20.26 \label{Tab:Q0012}}
\begin{tabular}{cccccccc}
\hline\hline
v	&b$_{eff}$	&	Mg I 			&	Mg II			&	Al II			&	Al III			&	Si II 			&		Fe II          		\\
\hline	
$-$61	&	5.3	&	-			&	$>$1.10E13		&	(1.16$\pm$0.14)E12	&	-			&	(4.08$\pm$2.27)E13	&	(1.66$\pm$0.18)E13	\\
$-$43	&	14.3	&	(4.40$\pm$0.32)E11	&	$>$6.38E13		&	$>$8.02E12		&	(2.23$\pm$0.15)E12	&	(1.65$\pm$0.41)E13	&	(1.48$\pm$0.05)E14	\\
$-$2	&	5.2	&	(9.32$\pm$1.97)E10	&	$>$3.23E13		&	$>$1.50E12		&	(3.79$\pm$0.16)E12	&	(4.05$\pm$1.44)E13	&	(4.75$\pm$0.56)E12	\\
57	&	13.7	&	-			&	$>$8.20E12		&	$>$8.57E11		&	(1.32$\pm$0.13)E12	&	(1.39$\pm$0.48)E13	&	(3.39$\pm$0.63)E12	\\
97	&	10.2	&	-			&	(2.63$\pm$0.11)E12	&	(2.03$\pm$0.57)E11	&	-			&		-		&	-			\\
127	&	12.7	&	-			&	(5.12$\pm$0.15)E12	&	(2.96$\pm$0.63)E11	&	(4.49$\pm$1.19)E11	&		-		&	(1.15$\pm$0.56)E12	\\
\hline
\end{tabular}
\end{minipage}
\end{table*}

\begin{table*}
\begin{minipage}{150mm}
\caption{Column densities for Q0021+0104, z$_{abs}$=1.3259, \nhI=20.04 \label{Tab:Q0021A}}
\begin{tabular}{ccccccc}
\hline\hline
v	&b$_{eff}$	&	Mg I			&	Mg II			&	Al II		&	Si II		&	Fe II			\\
\hline
$-$231	&	4.1	&	-			&	(2.00$\pm$0.21)E12	&	-		&	-		&	(2.53$\pm$0.90)E12	\\
$-$166	&	9.0	&	-			&	$>$1.06E13		&	$>$1.28E12	&	$>$1.07E13	&	(9.54$\pm$1.23)E12	\\
$-$127	&	13.0	&	-			&	$>$6.91E12		&	$>$2.04E12	&	-		&	(3.78$\pm$1.16)E12	\\
$-$101	&	8.4	&	-			&	$>$1.27E14		&	$>$5.96E12	&	$>$8.16E13	&	(3.42$\pm$0.25)E13	\\
$-$65	&	9.5	&	-			&	$>$2.31E13		&	$>$4.73E12	&	$>$3.54E13	&	(1.95$\pm$0.18)E13	\\
$-$35	&	13.8	&	(7.79$\pm$0.88)E11	&	$>$9.89E13		&	$>$1.81E13	&	$>$2.64E14	&	(1.06$\pm$0.06)E14	\\
$-$8	&	9.0	&	(6.60$\pm$0.83)E11	&	$>$1.32E14		&	$>$3.56E12	&	$>$8.19E13	&	(7.57$\pm$0.54)E13	\\
25	&	5.7	&	-			&	$>$4.01E13		&	$>$9.52E11	&	$>$2.84E13	&	(1.30$\pm$0.18)E13	\\
39	&	6.8	&	-			&	$>$1.04E13		&	$>$3.42E12	&	$>$6.79E13	&	(4.02$\pm$0.51)E13	\\
55	&	10.1	&	-			&	$>$1.73E14		&	$>$4.44E12	&	$>$9.53E13	&	(9.17$\pm$0.66)E13	\\
95	&	7.2	&	-			&	$>$7.12E13		&	$>$2.99E12	&	$>$6.42E13	&	(5.51$\pm$0.43)E13	\\
118	&	14.1	&	-			&	$>$1.82E13		&	$>$2.57E12	&	$>$4.97E13	&	(2.21$\pm$0.19)E13	\\
145	&	9.4	&	-			&	$>$6.21E12		&	-		&	$>$6.84E12	&	(1.29$\pm$0.14)E13	\\
\hline
\end{tabular}
\end{minipage}
\end{table*}

\begin{table*}
\begin{minipage}{150mm}
\caption{Column densities for Q0021+0104, z$_{abs}$=1.5756, \nhI=20.48 \label{Tab:Q0021B}}
\begin{tabular}{ccccccc}
\hline\hline
v	&b$_{eff}$	&	Mg I			&	Mg II			&	Al III			&	Si II			&	Fe II			\\
\hline
$-$339	&	13.6	&	-			&	(5.37$\pm$0.27)E12	&	-			&	(1.05$\pm$0.22)E13	&	(2.16$\pm$0.42)E12	\\
$-$305	&	8.6	&	(8.89$\pm$3.67)E10	&	(5.44$\pm$0.32)E12	&	-			&	(1.22$\pm$0.21)E13	&	(2.79$\pm$0.39)E12	\\
$-$264	&	6.6	&	-			&	(6.87$\pm$1.27)E11	&	-			&	-			&	-			\\
$-$197	&	5.5	&	-			&	(4.00$\pm$0.30)E12	&	-			&	(5.84$\pm$1.64)E12	&	(2.52$\pm$0.37)E12	\\
$-$145	&	12.1	&	-			&	(3.89$\pm$0.28)E12	&	-			&	(4.01$\pm$0.43)E13	&	(2.69$\pm$0.48)E12	\\
$-$133	&	4.3	&	(8.99$\pm$3.54)E10	&	(4.51$\pm$0.77)E12	&	-			&	(3.09$\pm$2.53)E12	&	(4.51$\pm$0.59)E12	\\
$-$113	&	9.2	&	(4.16$\pm$0.47)E11	&	$>$6.39E13		&	-			&	(1.24$\pm$0.20)E14	&	(2.40$\pm$0.14)E13	\\
$-$87	&	10.7	&	(2.79$\pm$0.45)E11	&	$>$3.11E13		&	-			&	(5.65$\pm$0.66)E13	&	(1.53$\pm$0.10)E13	\\
$-$63	&	10.1	&	(2.98$\pm$0.92)E11	&	$>$6.78E13		&	(1.82$\pm$0.32)E12	&	(8.27$\pm$2.27)E13	&	(6.77$\pm$1.05)E13	\\
$-$56	&	11.4	&	(2.17$\pm$1.02)E11	&	$>$1.47E13		&	-			&	(3.05$\pm$2.48)E13	&	(3.49$\pm$1.65)E13	\\
$-$37	&	13.3	&	(2.04$\pm$0.54)E11	&	$>$2.32E13		&	-			&	(8.57$\pm$0.98)E13	&	(1.17$\pm$0.08)E14	\\
$-$5	&	9.7	&	(2.95$\pm$0.44)E11	&	(6.90$\pm$0.39)E12	&	-			&	(1.79$\pm$0.25)E13	&	(1.19$\pm$0.07)E13	\\
37	&	9.1	&	(3.48$\pm$0.47)E11	&	$>$2.29E13		&	-			&	(4.39$\pm$0.55)E13	&	(7.04$\pm$0.69)E12	\\
54	&	10.6	&	(1.65$\pm$0.45)E11	&	$>$3.79E13		&	-			&	(5.56$\pm$0.70)E13	&	(2.72$\pm$0.17)E13	\\
81	&	11.3	&	(6.96$\pm$0.59)E11	&	$>$3.16E13		&	-			&	(8.74$\pm$1.06)E13	&	(3.96$\pm$0.23)E13	\\
106	&	11.8	&	(5.93$\pm$0.96)E11	&	$>$1.20E13		&	-			&	(3.57$\pm$1.02)E13	&	(5.25$\pm$2.29)E12	\\
115	&	12.6	&	(3.05$\pm$0.89)E11	&	$>$2.85E13		&	-			&	(5.33$\pm$0.87)E13	&	(3.89$\pm$0.27)E13	\\
152	&	9.2	&	-			&	(5.46$\pm$0.31)E12	&	-			&	(1.18$\pm$0.21)E13	&	(2.40$\pm$0.38)E12	\\
179	&	6.4	&	-			&	(7.33$\pm$1.29)E11	&	-			&	-			&	-			\\
272	&	4.4	&	-			&	(9.17$\pm$1.35)E11	&	-			&	-			&	-			\\
\hline
\end{tabular}
\end{minipage}
\end{table*}

\begin{table*}
\begin{minipage}{150mm}
\caption{Column densities for Q0427$-$1302, z$_{abs}$=1.4080, \nhI=19.04 \label{Tab:Q0427}}
\begin{tabular}{cccccc}
\hline\hline
v	&b$_{eff}$	&	Mg II			&    Al II			&	Si II		&	Fe II			\\
\hline
1	&	6.9	&	(5.27$\pm$0.31)E13	&	(1.13$\pm$0.07)E12	&(3.29$\pm$0.21)E13	&	(2.27$\pm$0.07)E13	\\
$-$50	&	8.3	&	(4.38$\pm$0.65)E11	&	-		 	&(3.35$\pm$0.87)E12	&	-			\\
$-$24	&	13.2	&	(1.21$\pm$0.09)E12	&	(4.37$\pm$0.61)E11	&(3.13$\pm$1.02)E12	&	-			\\

\hline
\end{tabular}
\end{minipage}
\end{table*}

\begin{table*}
\begin{minipage}{150mm}
\caption{Column densities for Q1631+1156, z$_{abs}$=0.9004, \nhI=19.70 \label{Tab:1631}}
\begin{tabular}{cccccc}
\hline\hline
v	&b$_{eff}$	&	Mg I			&	Mg II			&	Ca II			&	Fe II			\\
\hline
$-$67	&	11.1	&		-		&	$>$1.65E13		&	-			&	(1.18$\pm$0.06)E13	\\
$-$35	&	8.0	&	(4.39$\pm$1.39)E11	&	$>$2.69E13		&	(4.67$\pm$1.19)E11	&	(3.20$\pm$0.20)E13	\\
$-$20	&	8.5	&	(1.80$\pm$0.30)E12	&	$>$2.48E13		&	(4.27$\pm$1.19)E11	&	(6.24$\pm$0.43)E13	\\
6	&	7.6	&		-		&	$>$7.98E13		&	(4.70$\pm$1.16)E11	&	(2.13$\pm$0.13)E13	\\
56	&	8.0	&		-		&	(3.29$\pm$0.29)E12	&	-			&	(1.89$\pm$0.33)E12	\\
88	&	10.9	&		-		&	(2.29$\pm$0.24)E12	&	-			&	-			\\
\hline
\end{tabular}
\end{minipage}
\end{table*}

\begin{table*}
\begin{minipage}{150mm}
\caption{Column densities for Q2051+1950, z$_{abs}$=1.1157, \nhI=20.00 \label{Tab:Q2051}}
\begin{tabular}{ccccccccccccc}
\hline\hline
v	&b$_{eff}$	&	Mg  I			&	Mg II			&	Al II		&	Al III			&	Si II			\\	
\hline
$-$19	&	8.1	&	-			&	(2.95$\pm$0.17)E12	&	-		&	-			&	-			\\
1	&	10.6	&	(3.25$\pm$0.35)E11	&	$>$2.82E13		&	$>$9.81E12	&	(2.64$\pm$0.44)E12	&	-			\\	
25	&	9.6	&	(7.17$\pm$0.44)E11	&	$>$8.11E13		&	$>$9.09E12	&	(3.12$\pm$0.47)E12	&	-			\\
51	&	11.1	&	(9.24$\pm$0.60)E11	&	$>$7.42E13		&	$>$9.87E12	&	(4.18$\pm$0.65)E12	&	-			\\
66	&	8.0	&	(1.79$\pm$0.11)E12	&	$>$5.94E13		&	$>$8.01E12	&	(1.03$\pm$0.14)E13	&	(1.01$\pm$0.18)E15	\\
88	&	8.2	&	(4.00$\pm$0.37)E11	&	$>$8.17E13		&	$>$3.31E12	&	(5.86$\pm$0.74)E12	&	-			\\
105	&	7.4	&	(2.46$\pm$0.37)E11	&	$>$2.29E13		&	$>$3.62E12	&	(3.48$\pm$0.63)E12	&	-			\\
144	&	11.9	&	-			&	(2.17$\pm$0.13)E12	&	-		&	(8.78$\pm$3.70)E11	&	-			\\
119	&	9.8	&	-			&	$>$6.08E13		&	$>$6.04E12	&	(3.34$\pm$0.55)E12	&	(3.94$\pm$1.36)E14	\\
\hline
v	&b$_{eff}$	&	Ca II			&	Mn II			&	Cr II			&	Fe II			&	Zn II			\\
\hline
$-$19	&	8.1	&	 -			&	-			&	 	-		&	(2.68$\pm$0.71)E12	&	-			\\
1	&	10.6	&	-			&	(9.91$\pm$1.61)E11	&	 	-		&	(8.77$\pm$0.40)E13	&	(7.90$\pm$0.92)E11	\\
5	&	9.6	&	(7.05$\pm$1.29)E11	&	(2.27$\pm$0.18)E12	&	(2.46$\pm$1.15)E12	&	(1.59$\pm$0.09)E14	&	(1.34$\pm$0.09)E12	\\
51	&	11.1	&	(1.21$\pm$0.17)E12	&	(2.35$\pm$0.21)E12	&	 	-		&	(1.54$\pm$0.10)E14	&	(1.12$\pm$0.11)E12	\\
66	&	8.0	&	(8.85$\pm$1.51)E11	&	(4.88$\pm$0.27)E12	&	(5.37$\pm$1.30)E12	&	(2.81$\pm$0.34)E14	&	(2.59$\pm$0.12)E12	\\
8	&	8.2	&	(4.02$\pm$1.14)E11	&	(2.54$\pm$0.19)E12	&	 	-		&	(1.85$\pm$0.15)E14	&	(8.77$\pm$0.87)E11	\\
105	&	7.4	&	(3.49$\pm$1.18)E11	&	(1.55$\pm$0.19)E12	&	 	-		&	(1.06$\pm$0.09)E14	&	(3.84$\pm$0.90)E11	\\
144	&	11.9	&	-			&	(5.47$\pm$1.55)E11	&	 	-		&	(5.22$\pm$0.77)E12	&	-			\\
119	&	9.8	&	-			&	(9.87$\pm$1.77)E11	&	 	-		&	(5.83$\pm$0.32)E13	&	(8.01$\pm$1.00)E11	\\
\hline
\end{tabular}
\end{minipage}
\end{table*}

\begin{table*}
\begin{minipage}{150mm}
\caption{Column densities for Q2352$-$0028, z$_{abs}$=0.8739, \nhI=19.18 \label{Tab:Q2352A}}
\begin{tabular}{ccccccccccccc}
\hline\hline
v	&b$_{eff}$	&	Mg I			&	Mg II			&	Fe II			\\
\hline
$-$146	&	9.8	&	-			&	(1.32$\pm$0.15)E12	&	(5.18$\pm$3.11)E11	\\
$-$122	&	11.8	&	-			&	(4.03$\pm$0.23)E12	&	(1.80$\pm$0.35)E12	\\
$-$94	&	11.8	&	-			&	(1.62$\pm$0.16)E12	&	(1.10$\pm$0.34)E12	\\
$-$46	&	9.2	&	(1.91$\pm$0.58)E11	&	$>$1.41E13		&	(2.97$\pm$0.32)E12	\\
$-$15	&	9.0	&	(2.69$\pm$0.67)E11	&	$>$2.90E13		&	(8.25$\pm$0.41)E12	\\
0	&	7.5	&	(1.33$\pm$0.60)E11	&	$>$7.27E13		&	(6.27$\pm$0.38)E12	\\
26	&	6.4	&	-			&	$>$6.27E12		&	(2.30$\pm$0.39)E12	\\
34	&	7.7	&	-			&	$>$3.92E13		&	(1.62$\pm$0.41)E12	\\
58	&	5.6	&	(1.16$\pm$0.51)E11	&	$>$1.57E13		&	(4.03$\pm$0.30)E12	\\
79	&	7.4	&	-			&	$>$3.64E12		&	(1.30$\pm$0.29)E12	\\
\hline
\end{tabular}
\end{minipage}
\end{table*}

\begin{table*}
\begin{minipage}{150mm}
\caption{Column densities for Q2352$-$0028, z$_{abs}$=1.0318, \nhI=19.81 \label{Tab:Q2352B}}
\begin{tabular}{ccccccccccccc}
\hline\hline			 		  	
v	&b$_{eff}$	&	Mg I			&	Mg II			&	Al III			&	Si II			&	Cr II			&	Fe II			\\
\hline
$-$111	&	6.5	&	-			&	(1.23$\pm$0.15)E12	&	-			&	(1.36$\pm$0.52)E14	&	-			&	(2.26$\pm$0.44)E12	\\
$-$77	&	11.1	&	(8.43$\pm$0.54)E11	&	$>$2.54E14		&	(1.66$\pm$0.47)E12	&	(2.83$\pm$0.65)E14	&	-			&	(6.93$\pm$0.23)E13	\\
$-$53	&	9.4	&	(5.54$\pm$0.78)E11	&	$>$1.15E13		&	-			&	-			&	-			&	(7.02$\pm$0.55)E13	\\
$-$46	&	8.0	&	-			&	$>$2.83E13		&	(2.67$\pm$0.76)E12	&	(3.57$\pm$1.05)E14	&	-			&	(1.07$\pm$0.08)E14	\\
$-$28	&	7.6	&	(3.89$\pm$0.42)E11	&	$>$9.56E13		&	(3.01$\pm$0.49)E12	&	(2.49$\pm$0.60)E14	&	-			&	(4.41$\pm$0.20)E13	\\
0	&	8.8	&	(1.35$\pm$0.36)E11	&	$>$3.10E13		&	-			&	(2.04$\pm$0.61)E14	&	-			&	(4.99$\pm$0.21)E13	\\
18	&	10.2	&	(3.08$\pm$0.41)E11	&	$>$1.31E14		&	(3.83$\pm$0.53)E12	&	(4.52$\pm$0.70)E14	&	(2.12$\pm$0.77)E12	&	(1.20$\pm$0.04)E14	\\
48	&	6.4	&	(1.76$\pm$0.36)E11	&	$>$8.44E13		&	(2.04$\pm$0.44)E12	&	-			&	-			&	(4.24$\pm$0.21)E13	\\
66	&	8.9	&	(8.84$\pm$0.59)E11	&	$>$3.00E13		&	(6.25$\pm$0.62)E12	&	(5.99$\pm$0.72)E14	&	(4.07$\pm$0.77)E12	&	(1.45$\pm$0.06)E14	\\
88	&	8.4	&	(6.30$\pm$0.48)E11	&	$>$1.77E14		&	(6.11$\pm$0.60)E12	&	(4.81$\pm$0.67)E14	&	(2.88$\pm$0.72)E12	&	(1.51$\pm$0.06)E14	\\
111	&	4.0	&	(9.74$\pm$2.96)E10	&	$>$1.92E13		&	-			&	-			&	-			&	(4.71$\pm$0.47)E12	\\
\hline
\end{tabular}
\end{minipage}
\end{table*}

\begin{table*}
\begin{minipage}{150mm}
\caption{Column densities for Q2352$-$0028, z$_{abs}$=1.2467, \nhI=19.60 \label{Tab:Q2352C}}
\begin{tabular}{ccccccccccccc}
\hline\hline			 	
v	&b$_{eff}$	&	Mg I			&	Mg II		&	Al III			&	Fe II			\\
\hline
$-$210	&	9.5	&	-			&	$>$5.33E12	&	-			&	(3.37$\pm$0.74)E12	\\
$-$184	&	10.9	&	-			&	$>$1.98E12	&	-			&	-			\\
$-$152	&	7.9	&	(1.49$\pm$0.37)E11	&	$>$3.60E13	&	-			&	(5.71$\pm$0.77)E12	\\
$-$109	&	12.7	&	-			&	$>$4.54E12	&	-			&	-			\\
$-$87	&	8.9	&	-			&	$>$6.93E12	&	(1.41$\pm$0.25)E12	&	(1.79$\pm$0.78)E12	\\
$-$66	&	9.3	&	(2.73$\pm$0.43)E11	&	$>$2.03E14	&	(7.92$\pm$0.53)E12	&	(4.63$\pm$0.23)E13	\\
$-$31	&	8.9	&	-			&	$>$1.44E14	&	(4.27$\pm$0.34)E12	&	(8.28$\pm$0.85)E12	\\
6	&	7.8	&	(2.88$\pm$0.44)E11	&	$>$2.15E14	&	(3.95$\pm$0.34)E12	&	(1.12$\pm$0.10)E13	\\
21	&	5.7	&	(5.34$\pm$0.56)E11	&	$>$1.80E14	&	(1.69$\pm$0.25)E12	&	(2.03$\pm$0.15)E13	\\
50	&	4.0	&	(2.40$\pm$0.41)E11	&	$>$5.91E14	&	(5.28$\pm$1.75)E11	&	(8.61$\pm$0.96)E12	\\
77	&	6.6	&	-			&	$>$1.64E12	&	-			&	-			\\
108	&	11.6	&	(1.84$\pm$0.41)E11	&	$>$1.60E13	&	(1.99$\pm$0.27)E12	&	(5.54$\pm$0.82)E12	\\
135	&	4.5	&	-			&	$>$2.78E12	&	-			&	(2.79$\pm$0.67)E12	\\
163	&	10.5	&	(4.83$\pm$0.48)E11	&	$>$3.40E14	&	(2.53$\pm$0.28)E12	&	(4.93$\pm$0.22)E13	\\
202	&	7.9	&	-			&	$>$2.19E12	&	-			&	-			\\
\hline
\end{tabular}
\end{minipage}
\end{table*}

\label{lastpage}

\end{document}